%% file: main.tex
\documentclass[conference]{IEEEtran}
\usepackage{url} 
\usepackage{graphicx}
\usepackage{tikz}
\usepackage{amsmath}
\usepackage{amsthm}
\usepackage{bbm}
\usepackage{booktabs}
\usepackage{paralist}
\usepackage{cleveref}
\usepackage[frozencache,cachedir=minted-cache]{minted}
\usepackage{listings}
\usepackage{color}
\usepackage{comment}
\usepackage{subcaption}
\usepackage{xspace}
\usepackage{adjustbox}
\usepackage{tikz}
\usepackage{pgfplots}
\usepackage{pgfplotstable}
\pgfplotsset{compat=1.18}
\usepackage{authblk}
\usepackage{siunitx}
\usepackage{thmtools,thm-restate}

\definecolor{dkgreen}{rgb}{0,0.6,0}
\definecolor{gray}{rgb}{0.5,0.5,0.5}
\definecolor{mauve}{rgb}{0.58,0,0.82}

\newcommand*\emptycirc[1][1ex]{\tikz\draw[thick] (0,0) circle (#1);} 
\newcommand*\halfcirc[1][1ex]{%
  \begin{tikzpicture}
  \draw[fill] (0,0)-- (90:#1) arc (90:270:#1) -- cycle ;
  \draw[thick] (0,0) circle (#1);
  \end{tikzpicture}}
\newcommand*\fullcirc[1][1ex]{\tikz\fill (0,0) circle (#1);} 

\newtheorem{definition}{Definition}
\newtheorem{example}{Example}

\newcommand{\advantage}{\mathsf{Adv}}
\newcommand{\sidechannels}{\mathsf{SC}}
\newcommand{\attackeroutput}{\mathsf{output}}
\newcommand{\attacker}{\mathcal{A}}
\newcommand{\rv}[1]{\mathsf{#1}}
\newcommand{\para}[1]{\smallskip\noindent\textbf{{#1}}}

\usepackage{tikz}
\usepackage{letltxmacro}

\makeatletter

\def\FrameWorkName{\mbox{SNPeek\hspace{-0.08em}}\xspace}

\title{\FrameWorkName: Side-Channel Analysis for Privacy Applications on Confidential VMs}

\newif\ifanon

\ifanon
    \author{Anonymous Submission}
\else 
    \author[1,2]{Ruiyi Zhang}
    \author[2]{Albert Cheu}
    \author[2]{Adria Gascon}
    \author[2]{Daniel Moghimi} 
    \author[2]{\\Phillipp Schoppmann}
    \author[1]{Michael Schwarz}
    \author[2]{Octavian Suciu}
    \affil[1]{CISPA Helmholtz Center for Information Security}
    \affil[2]{Google}
\fi 

\begin{document}

\IEEEoverridecommandlockouts
\makeatletter\def\@IEEEpubidpullup{6.5\baselineskip}\makeatother
\IEEEpubid{\parbox{\columnwidth}{
		Network and Distributed System Security (NDSS) Symposium 2026\\
		23 - 27 February 2026 , San Diego, CA, USA\\
		ISBN 979-8-9919276-8-0\\  
		https://dx.doi.org/10.14722/ndss.2026.240699\\
		www.ndss-symposium.org
}
\hspace{\columnsep}\makebox[\columnwidth]{}}

\maketitle

\input{00.abstract}

\input{01.intro}
\input{02.prelim}
\input{03.threatmodel}
\input{04.framework}
\input{06.pir}

\input{07.phh}
\input{08.udf} 
\input{10.related}
\input{99.conclusion}
\input{99a.ethics}

\section*{Acknowledgements}
We would like to thank Kobbi Nissim, Jonathan Katz, Sarah Meiklejohn, Marco Gruteser, Peter Kairouz, Daniel Ramage and Shabsi Walfish for their constructive feedback and support. 

\bibliographystyle{IEEEtran}
{\footnotesize
\bibliography{references}
}
\appendix
\input{100.appendices}

\end{document}

%% file: 00.abstract.tex
\begin{abstract}
Confidential virtual machines (CVMs) based on trusted execution environments (TEEs) enable new privacy-preserving solutions.
Yet, they leave side-channel leakage outside their threat model, shifting the responsibility of mitigating such attacks to developers.
However, mitigations are either not generic or too slow for practical use, and developers currently lack a systematic, efficient way to measure and compare leakage across real-world deployments.

In this paper, we present \FrameWorkName, an open-source toolkit that offers configurable side-channel tracing primitives on production AMD SEV-SNP hardware and couples them with statistical and machine-learning-based analysis pipelines for automated leakage estimation.
We apply \FrameWorkName\ to three representative workloads that are deployed on CVMs to enhance user privacy---private information retrieval, private heavy hitters, and Wasm user-defined functions---and uncover previously unnoticed leaks, including a covert channel that exfiltrated data at 497 kbit/s.
The results show that \FrameWorkName\ pinpoints vulnerabilities and guides low-overhead mitigations based on oblivious memory and differential privacy, giving practitioners a practical path to deploy CVMs with meaningful confidentiality guarantees.
\end{abstract}

%% file: 01.intro.tex
\section{Introduction}
Cloud providers now offer \emph{confidential virtual machines} (CVMs) based on hardware architectures such as AMD SEV-SNP, and Intel TDX.
These CVMs encrypt guest memory and enforces security via the hardware, allowing a tenant to run unmodified binaries while keeping data hidden from the hypervisor and other co-tenants.
Unfortunately, Intel, AMD, and ARM (e.g., CCA) \emph{explicitly exclude} leakage through page-table activity and processor-cache state from their CVM's threat model.
Therefore, 
such side-channel attacks based on page tables~\cite{xu2015controlled} and caches~\cite{liu2015last,moghimi2017cachezoom} can track memory accesses at 4\,kB and 64\,B granularities, respectively.
They have been very successful at, e.g., inferring cryptographic keys~\cite{aranha2020ladderleak,lyu2018survey}.
That setting matches the traditional notion of (architectural) side-channel attack, where the adversary's goal is to extract a private key, e.g., a signing key, from a confidential VM by exploiting or inducing side-channel leakage.
While implementing successful mitigations has proven challenging, best practices such as constant-time code are well understood for {\em concrete} cryptographic applications, e.g., RSA-based signature schemes. This is the result of a fruitful line of security research that 
provided a feedback mechanism to chip manufacturers.

\para{From cryptographic applications to privacy-preserving data analyses.}
A recent trend in industry involves adopting CVMs for user data processing, e.g., computing user statistics, distributed secure computation, and oblivious data retrieval~\cite{li2022sok,eichner2024confidential, tfblogpost, srinivas2024federated}. Just like with cryptographic code, these data-driven applications 
inherit a large attack surface when deployed in CVMs~\cite{conffed,privacysandbox}, and mitigations are 
delegated to the application developers.

Privacy threats in these data-driven applications are less precisely defined and more challenging to quantify compared to those in cryptographic applications. An attacker clearly defeats a cryptographic application when they recover a pseudorandom key. In contrast, when an attacker recovers an input to a data-processing application, their success must be evaluated in light of both their prior knowledge and the goals of the application itself. Without \textbf{a rigorous formalization of the threat model and attack success}, it will be tempting to fall back to impractical mitigations like constant-time code for \emph{general} software~\cite{yuan2022automated,shahverdi2021database,yan2020cache,wang2022stealthy}, instead of optimizing defenses based on the specific workload. Aside from a strong definition, practitioners also need a \textbf{systematic way to quantify, measure, compare, and reduce the leakage of real workloads in the deployed scenario}.

In this paper, we provide formal definitions and an evaluation framework called \textbf{\FrameWorkName}. Our goal is to enable the systematic investigation of side-channel leakage in privacy-preserving workloads, and potential mitigations.

\para{Quantifying Attacker's Success (Section~\ref{sec:threat_model}).} Taking inspiration from the literature on cryptography and differential privacy, we define and measure success of privacy attacks in a relative sense: assuming the attacker has some prior knowledge about a target's data, we measure success by comparing the attacker’s probability of correctly guessing the data before and after the attack.
If the prior probability of a successful guess is already high, our formulation captures the intuition that there is not much more information an attacker can learn.
Additionally, our model accounts for the fact that the party connecting the CVM to the outside world can introduce \emph{Sybils}, i.e., arbitrarily well-crafted values that can trigger more leakage than naturally-occurring inputs.

\para{Automated Side-channel Extraction and Analysis (Section~\ref{sec:framework}).} Our open-source toolkit, 
currently implemented for AMD SEV-SNP, consists of a \emph{trace extraction} and a \emph{trace analysis} phase.
In the former, \FrameWorkName\ records low-noise page-table~\cite{xu2015controlled,van2017telling,wang2017leaky} and cache traces~\cite{percival2005cache,liu2015last,moghimi2017cachezoom}, and optionally also performance-counter values~\cite{gast2025counterseveillance,uhsadel2008exploiting} and ciphertexts~\cite{li2021cipherleaks,li2022systematic}, from commodity SEV-SNP guests.
We introduce a noise-free and efficient Multi-Prime+Probe attack targeting 64 cache sets on AMD CPUs with a non-inclusive last-level cache by exploiting model-specific registers that allow restriction of the L3 cache. 
Our optimizations significantly reduce measurement noise and improve the practicality of the cache attack.
Additionally, we devise filtering strategies that restrict trace collection to the relevant part of the application, thereby speeding up measurement.
In the analysis phase, \FrameWorkName\ analyzes those traces with a set of predefined statistics and machine-learning models for automated side-channel traces analysis.
These predefined models allow for easy pinpointing of the leakage source and experimentation with attackers of various capabilities.

\para{Evaluation on Real-World workloads (Sections~\ref{sec:pir},~\ref{sec:phh},~\ref{sec:udf}).} We apply \FrameWorkName\ to evaluate side-channel leakage of three real-world privacy applications that are executed on top of CVMs: Private Information Retrieval (PIR), Private Heavy Hitters (PHH), and User Defined Functions (UDF).
In PIR, a party wishes to retrieve an element from a remote database without letting the database's maintainer(s) learn which element was accessed.
TEE-based PIR implementations are available as open-source projects such as Project Oak~\cite{oak} and the Signal messenger~\cite{signal-blog}.
We analyze the effectiveness of mitigations such as ORAM~\cite{StefanovDSCFRYD18,signal-blog} and demonstrate how our framework can uncover and pinpoint subtle leakage.
Surprisingly, we show that even constant-time ORAM may exhibit leakage when deployed on AMD SEV-SNP due to ciphertext side channel leakage.

We additionally demonstrate how \FrameWorkName\ can evaluate the privacy guarantee of a PHH application as implemented by the TensorFlow Federated project \cite{fedTF} and deployed by Google~\cite{tfblogpost}.
In this application, a large number of personal devices (e.g., smartphones) hold sensitive data, such as location or browser history, and the service provider wishes to identify frequently occurring entries in this distributed data store
in a differentially private manner.
We show that the Tensorflow-Federated implementation~\cite{fedTFphh}, which is not leakage-aware, is vulnerable to a privacy attack due to its data-dependent execution behavior when deployed on AMD SEV-SNP.
We present a series of examples to illustrate the use of our framework to detect issues, develop and evaluate defenses, and advanced attacks. 
Along the way, we also introduce a partial mitigation based on differential privacy that might be of independent interest.

Finally, we demonstrate how \FrameWorkName can evaluate private user-defined functions.
In-memory data stores such as those used in PIR or PHH may also support custom queries via a user-defined function (UDF)~\cite{aksu2023summary,privacysandbox_fledge}.
For example, in the context of Google's Privacy Sandbox~\cite{privacysandbox}, UDFs based on the Wasm language are written by AdTechs to customize higher-level (privacy-preserving) aggregations about end-users' web browsing behavior~\cite{privacysandbox_udf,oak}.
Private UDFs introduce an attack scenario where the attacker can not only collect side-channel traces outside the CVM, but also introduce new queries on processed data sources and efficiently steal data via a covert channel.
Our results show that a covert-channel attack can leak data from a UDF inside the Wasm language runtime~\cite{privacysandbox_fledge} to a colluding hypervisor at a rate of at least 497 kbit/s.

\noindent\textbf{Contributions.}
We summarize our contributions as follows.
\begin{compactenum}
  \item We introduce \FrameWorkName, a modular framework that gathers various side-channel traces, including a novel noise-resilient Prime+Probe, 
  from unmodified AMD SEV-SNP guests and provides different trace filters to reduce collection overhead. 
  The framework includes statistical and machine-learning models to automatically analyze gathered side-channel traces to enable leakage estimates for non-domain experts.
  \item We introduce and motivate rigorous quantitative notions of privacy leakage via side-channel in the presence of a malicious attacker, i.e., the attacker's advantage. Our notion is inspired by the privacy attacks literature and can be easily estimated empirically using \FrameWorkName's ML components. 
  \item We evaluate three representative privacy workloads---PIR, private heavy hitters (PHH), and user-defined functions (UDFs) based on Wasm---and show how \FrameWorkName\ guides the design of effective, low-overhead mitigations. 
\end{compactenum}
Although our toolkit is currently implemented on AMD SEV-SNP, we anticipate many of our attacks carry over to other vendors with minimal changes. 

\para{Responsible Disclosure.}
Our research follows established responsible disclosure guidelines. 
We notified maintainers of all open-source projects whose applications showed vulnerabilities under our framework---specifically Project Oak~\cite{oak}, TensorFlow Federated~\cite{fedTF}, and the Privacy Sandbox~\cite{privacysandbox_fledge}. 
Each project acknowledged the security impact of software-based side-channel attacks and indicated ongoing work to strengthen its privacy protections.

\para{Availability.}
The source code of \FrameWorkName\ is open-sourced at \url{https://github.com/google-parfait/cvm-side-channel-analysis}.

%% file: 02.prelim.tex
\section{Preliminaries}

\subsection{Confidential VMs}
Confidential VMs are based on hardware-based trusted execution environments, such as AMD SEV-SNP~\cite{sev2020strengthening} or Intel TDX~\cite{tdx}.
They rely on hardware-based access control and memory encryption to prevent other VMs and privileged software (hypervisor, BIOS) from accessing the memory of a trusted domain (a CVM instance).
Additionally, memory is encrypted as soon as it leaves the CPU. 
The operating system is only responsible for the availability of trusted workloads (e.g., scheduling workloads, mapping memory, and handling I/O), and can thus be untrusted.

CVMs additionally rely on a hardware-based root of trust (external to the CPU core) and a remote-attestation protocol to guarantee the integrity of the software and hardware components responsible for executing a trusted domain.
Therefore, before a user sends encrypted data to the CVM, they can verify that the data is processed by genuine hardware and the right software components, including the latest firmware and microcode security patches.

\subsection{Software-Based Side Channels}
Software-based side-channel attacks exploit shared resources and exposed system interfaces to leak information about computation of other users on the system.
Some of the attack primitives that are relevant to our work focusing on CVMs include:

\para{Page table.}
Controlled-channel attacks target page tables, allowing a malicious hypervisor to track a program's memory access pattern at page-level granularity (typically 4\,kB).
This attack vector applies to various CVM platforms.
For instance, a hypervisor can unmap a guest memory page on AMD SEV-SNP.
While Intel TDX (and ARM CCA) restricts such direct page-table  manipulation, similar leakage is achievable: privileged software can leverage the TDX module to block and unblock memory ranges, resulting in a similar VM exit as soon as a trusted domain accesses the memory ranges~\cite{xu2015controlled}.
%

\para{Cache.}
Cache attacks, such as the Prime+Probe technique~\cite{osvik2006cache}, target shared caches in modern CPUs.
These remain applicable to CVM platforms like AMD SEV-SNP, Intel TDX, and ARM CCA, given their reliance on shared-cache architectures.
In a Prime+Probe attack, the attacker fills a cache set with known addresses and waits for the victim to access data mapping to the same cache set.
After the victim's access, the attacker detects which parts have been evicted by measuring timing differences of re-accessing its own addresses.

\para{HPC.}
Hardware performance counters (HPCs) are special registers in modern CPUs that track various microarchitectural events, such as cache hits, misses, and branch predictions.
Privileged attackers can use performance counters to gather detailed information to infer sensitive information, such as cryptographic keys or execution-flow patterns~\cite{gast2025counterseveillance}.

\para{Ciphertext visibility.}
Ciphertext side-channel attacks~\cite{li2021cipherleaks} exploit the memory encryption scheme in AMD SEV-SNP.
Each 16-byte-aligned memory block is encrypted individually, using a tweak value derived from its physical address.
At a specific address, the same plaintext always produces the same ciphertext.
Although SEV-SNP aims to provide confidentiality and integrity, a malicious hypervisor can read the encrypted memory.
By observing changes in ciphertexts, the attacker can infer changes in the underlying plaintexts, beyond learning that a given region in memory did change.

\input{mitigations}

\subsection{Differential Privacy (DP)}

Suppose there are $n$ distinct inputs $X=X_1,\dots,X_n$ to a computational service $S$, where each input $X_i$ may be a sensitive value (i.e., proprietary information or personal attribute) of a distinct input provider. Let $V^A_S(X)$ denote attacker $A$'s view of that service when $X$ is given as input. In the textbook \emph{central model}, the view is the output of the service, like an estimate of a mean or a table of synthetic data \cite{dmns06}. 
\newcommand{\pr}[2]{{\ifx&#1& \mathbbm{P} \else \underset{#1}{\mathbbm{P}} \fi \left[#2\right]}}
$S$ ensures $(\varepsilon,\delta)$-DP against $A$ if, for any $X,X'$ that differ on any one input and any possible $Y$, $\pr{}{V^A_S(X) \in Y} \leq e^\varepsilon \cdot \pr{}{V^A_S(X') \in Y}+\delta$.

We emphasize that the guarantee must hold for \emph{all} neighboring inputs $X,X'$. This effectively means that the attacker has narrowed down a target's input to one of two different values and controls all other inputs (\emph{Sybils}).

There is considerable risk in underestimating the scope of an adversary's view. As highlighted in prior work~\cite{HaeberlenPN11,ratliff-vadhan}, consider a service that outputs the same mean estimate on two neighboring datasets but whose running time differs dramatically: an adversary can deduce the target's input whenever it can measure elapsed time. Haeberlen~et~al.~\cite{HaeberlenPN11} attempt to mitigate this by modeling the adversary as only able to access $S$ via a network connection and a restricted query language. Meanwhile, Ratliff and Vadhan~\cite{ratliff-vadhan} carefully reason about sensitivity and inject random-length delays, following the pattern of the Laplace and Gaussian mechanisms. We note that our definition of DP is a strict generalization of the one by Ratliff and Vadhan~\cite{ratliff-vadhan}, since our adversary's view can encompass more than the timing side channel (e.g., memory access patterns).

%% file: mitigations.tex
\newcommand*\emptycircle[1][.7ex]{\tikz\draw[thick] (0,0) circle (#1);} 
\newcommand*\halfcircle[1][.7ex]{%
  \begin{tikzpicture}
  \draw[fill] (0,0)-- (90:#1) arc (90:270:#1) -- cycle ;
  \draw[thick] (0,0) circle (#1);
  \end{tikzpicture}}
\newcommand*\dottedcircle[1][1ex]{%
  \begin{tikzpicture}
    \foreach \x/\y in {-0.3/0.3, 0/0.3, 0.3/0.3, -0.3/0, 0/0, 0.3/0, -0.3/-0.3, 0/-0.3, 0.3/-0.3} {
      \fill (\x*#1,\y*#1) circle (0.1*#1);
    }
    \draw[thick] (0,0) circle (#1);
  \end{tikzpicture}}
\newcommand*\fullcircle[1][.7ex]{\tikz\fill (0,0) circle (#1);}


\begin{table}[h]
\centering
\caption{\fullcirc~: vulnerable, \emptycirc~: mitigated, \halfcirc~: partially, \\ \dottedcircle~: mitigation planned
}

\adjustbox{max width=0.8\linewidth}{
\begin{tabular}{lllll}
\toprule
\textbf{Platform} & \textbf{Page-level} & \textbf{Cache-level} & \textbf{Ciphertext} & \textbf{HPC} \\
\midrule
SEV-SNP Zen3/4      & \fullcirc                 & \fullcirc           & \fullcirc                & \fullcirc                                                      \\
SEV-SNP Zen5      & \fullcirc                 & \fullcirc           & \dottedcircle~\cite{Kalra2024cipher}                & \dottedcircle~\cite{amczen5hpc}                                                       \\
TDX         & \fullcirc~\cite{aktas2023intel}                 & \fullcirc~\cite{aktas2023intel,wilke2024tdxdown}          & \emptycirc                 & \halfcirc~\cite{tdxHpc}                                                       \\
CCA           & \fullcirc                 & \fullcirc           & \emptycirc                 & \halfcirc                              \\
\bottomrule
\end{tabular}
}
\label{tab:sysmit}
\end{table}

\para{System-level mitigations.} 
%
\Cref{tab:sysmit} presents the current state of system-level mitigation for our attack vectors.
Ciphertext side-channel attacks~\cite{li2021cipherleaks} are specific to SEV-SNP, and
AMD has provided software workarounds that make constant-time code even harder to implement~\cite{ctextSEV}.
%
Intel TDX and ARM CCA prevent the hypervisor from accessing guest-encrypted memory~\cite{aktas2023intel}.
%
AMD plans to address the leaks from ciphertext and HPCs on Zen 5 processors using ciphertext hiding~\cite{Kalra2024cipher} and PMC virtualization~\cite{amczen5hpc}, respectively. 
Currently, neither is supported in KVM.
%
For ARM CCA, which defines a broader architecture, performance monitoring virtualization for the trusted realm is platform dependent. 
However, cache attacks and page-level leakage remain unmitigated across SEV~\cite{sev2020strengthening}, TDX~\cite{aktas2023intel,wilke2024tdxdown}, and CCA~\cite{arm2021cca}.
They are considered out of scope by the vendors.
Thus, from the vendor's perspective, attacks on a given workload leveraging these side-channels are the responsibility of the application developer.





%% file: 03.threatmodel.tex
\section{Defining Side-Channel Privacy Attacks}
\label{sec:threat_model}
\begin{figure}
    \centering
    \includegraphics[width=.87\columnwidth]{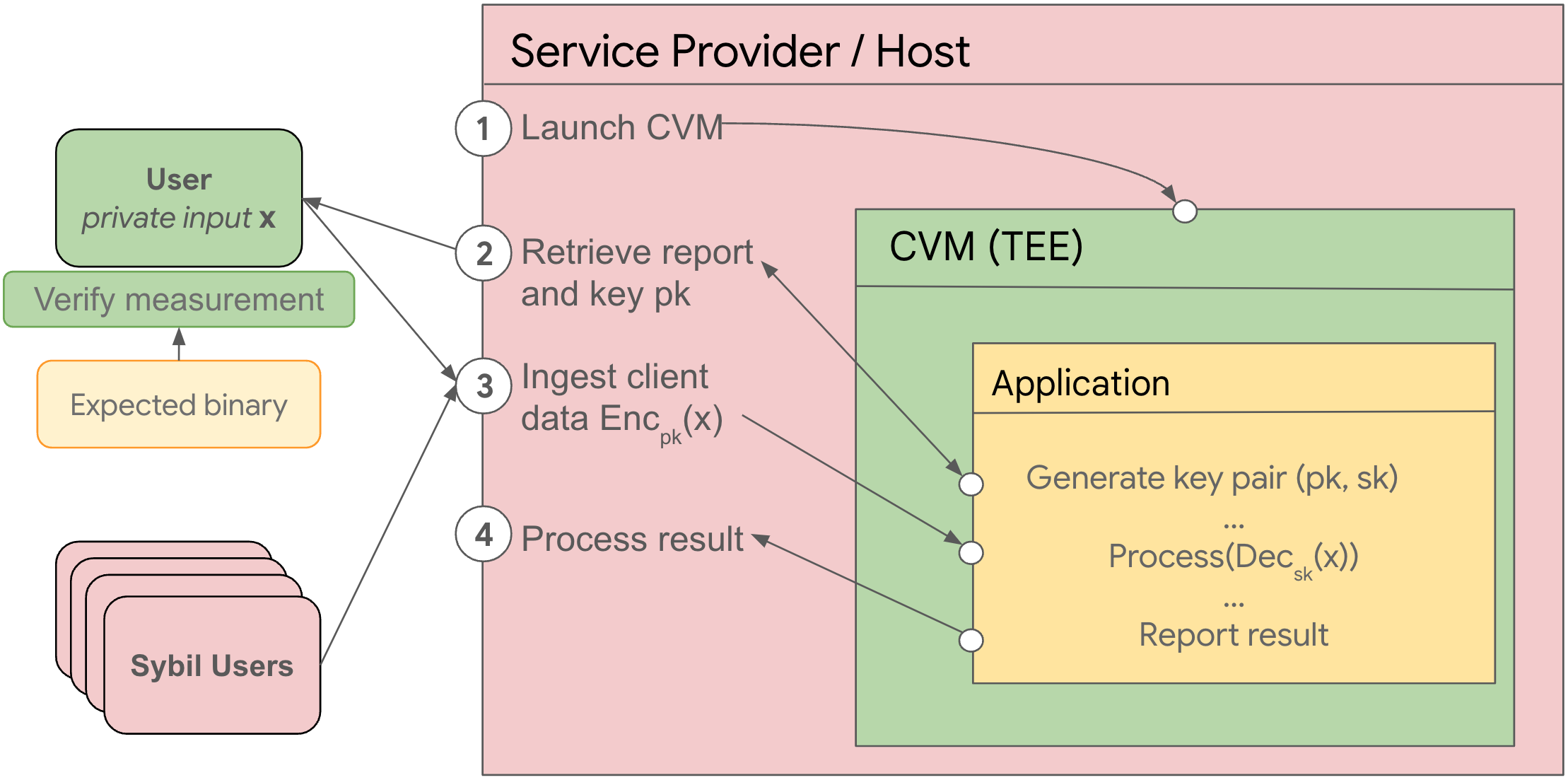}
    \caption{The user requests an attestation report and a public key before sending encrypted data to the isolated CVM. 
    The service provider and other users are untrusted. The CVM and application binary is trusted, but it may have the ability to execute user-defined queries from an untrusted source.} 
    \label{fig:threat-model}
\end{figure}

Here, we formalize how an attacker in the CVM threat model can recover information about sensitive inputs using side channels \cite{coco}. This includes personal attributes that are processed by software running in the CVM (e.g., location, webpage visits), not just cryptographic keys.

We assume a shared cloud environment where the hypervisor and other VMs are untrusted.
Figure~\ref{fig:threat-model} shows the structure
of a CVM-based data analysis system.
Data and/or custom queries are ingested from \emph{input providers}, also known as \emph{users}.
An untrusted \emph{service provider}---the party operating the data analysis service---seeks to reconstruct more information about the inputs than what is contained in the output. 
We use the term \emph{attacker} as a short synonym for the service provider. 
We assume no collusion between the service provider and the hardware manufacturer.
In Section~\ref{sec:udf}, we extend the model to assume that the attacker can provide custom queries to the protected key-value service.

{\em Sybils}~\cite{sybilattack} are a significant feature in our model. We do not assume public key infrastructure free from the influence of the service provider, which means the attacker can generate fake identities to take over a service: they can inject maliciously-generated inputs, and suppress honest inputs. 
Sybil attacks in distributed data analysis are a known issue, particularly in the context of federated learning~\cite{eichner2024confidential}. The attacks we design reaffirm this deep challenge.
However, we acknowledge a practical limitation against hardened systems using third-party privacy gateways~\cite{apple_requestflow}. To guarantee a victim's input is routed to a compromised machine for side-channel analysis, an \emph{attacker} must collude with the gateway, as monitoring all endpoints is undesirable. This countermeasure is thus effective against \emph{attackers} when the external party gateway remains uncompromised.

Like previous work on software-based attacks, we leave physical attacks out of the scope \cite{badramsp25,chen2021voltpillager}, assuming appropriate physical security is in place. 
Likewise, we exclude CPU bugs such as transient-execution attacks~\cite{Vanbulck2018foreshadow,Schwarz2019ZL,VanSchaik2019RIDL,moghimi2023downfall} and CacheWarp~\cite{zhang2024cachewarp}, and software-based fault attacks like Rowhammer~\cite{Kim2014} and Plundervolt~\cite{Murdock2019plundervolt}.
We also assume that all applications are protected against rollback attacks, and therefore, honest client contributions cannot be duplicated or replayed by the attacker without aborting the application.

We sketch the steps of a generic data-analysis service by breaking it into an \emph{offline} and \emph{online} phase. 
The offline phase describes what happens before any interactions.

\noindent
\textbf{Offline phase}:
\begin{compactenum}
    \item \underline{Input Preparation}: The input providers generate their inputs. 
    The attacker does not know any target's input with complete certainty, but might have some prior knowledge; we model that input as being drawn from a probability distribution known to the attacker.

    \item \underline{Setup}: The guest VM binary is made reproducible for the purposes of verifiability. 
    Here, an attacker has full control to assess the behavior of the binary on their TEE-supporting hardware (e.g., debug mode) but does not have access to the (secret) data. In particular, the attacker can obtain statistical information to characterize secret inputs given side-channel information, either by ``manual'' inspection, or by training ML models.
    
\end{compactenum}
\noindent
\textbf{Online phase}:
\begin{compactenum}    
    \item \underline{Launch of CVM}: The service provider triggers the initialization of the trusted environment.
    
    \item \underline{Establishing trust}: The CVM's attestation report and public key are forwarded by the service provider to the users, who validate the report.
    
    \item \underline{Input Ingestion}: The service provider forwards a stream of encrypted user data and/or user-defined functions for the CVMs. Here,
    \begin{compactenum}
        \item the attacker can drop honest inputs and insert \emph{Sybils}, specially-crafted inputs that are meant to trigger more side-channel leakage.  However, we assume they do not duplicate or replay honest inputs.
        \item the attacker monitors side channels of the ingestion computation.
    \end{compactenum} 
    
    \item \underline{Report}: the guest VM computes a plaintext output that the service provider relays to its recipient. The attacker monitors side channels of the report computation.
\end{compactenum}

This gives a high-level idea of the actions an attacker can perform. Next, we provide a notation for the attacker's knowledge and define what it means for an attack to succeed.

\subsection{Attacker Knowledge \& Success}

The attacker is not necessarily limited to the knowledge gleaned from the service's execution: 
they may have \emph{prior information} about a target input provider. For example, they may know Alice is contributing URLs as input and that she only reads English. 
This means the attacker can rule out strings that are not URLs while also weighing URLs with a ``jp'' extension as less likely to have been visited by Alice. We will use $W$ to denote the probability distribution that describes this prior knowledge about a target. 

Although it is tempting to use the likelihood of reconstructing secrets (e.g., pseudorandom keys) as the metric of success, this is not always the correct choice:
if the prior $W$ is sufficiently skewed, the attacker may have a high probability of a correct guess \emph{even without looking at any side channels or CVM output}. In the case of URL visits, they are \textbf{not} uniformly random: if \texttt{example.com} is known to be $p$\% of all visits, then the baseline ``attack'' that simply outputs \texttt{example.com} has a $p$\% chance of being correct for a target. Thus, a high probability of a correct guess could be due to using leakage or reporting an argmax of a heavily-skewed $W$. As such, it is a poor metric to gauge success.

We take the stance that the quantity to measure is the \emph{advantage} afforded by the attack over the baseline argmax-of-$W$ strategy, the improvement in the probability of a correct guess\footnote{For cryptographic keys, the baseline is close to zero, so the advantage of an attack is close to the probability of reconstructing keys.}. We note that $\Omega(1)$ advantage is permissible in some applications; for example, DP computations already grant a ``privacy budget'' $\varepsilon$ and we show $\varepsilon=\Omega(1)$ permits $\Omega(1)$ advantage (see next section). Otherwise, the advantage should be bounded by a negligible function.

\subsection{Pairwise Distinguishability Attacks}
\label{sec:threat_model_distinguishability}

Let us consider a specific class of prior $W$: those distributions over two values that give equal probability to each. This occurs when the attacker is evenly split between, say, whether the last URL entered by a target individual was \texttt{example1.com} or \texttt{example2.com}. Without the side-channel leakage, the baseline argmax-of-$W$ strategy results in a guess that is right $1/2$ of the time. With the side-channel leakage, we would like to bound the \emph{advantage} over that baseline chance. The adversary observes these side channels via the pairwise distinguishability game:

\begin{definition}[Pairwise Distinguishability Game]\label{def:dist-game}
Let $\attacker$ be an attacker, and let $B$ be a binary executed in a CVM. Let $\sidechannels_\attacker^B(D)$ denote the leakage function for $\attacker$ when running $B$
on input $D$. Assuming uniform prior $W$ over $\{x_0, x_1\}$, the distinguishability game proceeds as follows:

\begin{compactenum}
    \item (Input Preparation) $\rv{c}$ is chosen randomly from $\{0, 1\}$, such that user's data $x_\rv{c}$ is $x_0$ or $x_1$ with a 1/2 chance
    \item (Before Input Ingestion) Attacker chooses Sybils $X$
    \item (Input Ingestion) Whole dataset $D$ is formed by appending $x_\rv{c}$ to $X$
    \item (After Report) $\attackeroutput_\attacker := \attacker(\sidechannels_\attacker^B(D))$
\end{compactenum}

\end{definition}
The distinguishability advantage of attacker $\attacker$
is $\advantage_\attacker := \max(0 , \Pr[\attackeroutput_\attacker = \rv{c}] - 0.5)$. 
The maximum value for this is $0.5$; we compute a \emph{normalized advantage} $\advantage_\attacker/(0.5)$ that ranges from 0 to 1 (least to most successful attack). Note that we can derive a bound on this advantage when leakage $\sidechannels_\attacker^B(D)$ satisfies DP:

\begin{restatable}[DP bounds advantage]{lem}{dpAdvantageLemma}
\label{lem:distinguishability-advantage}
If the leakage of $B$ guarantees $(\varepsilon,\delta)$-DP, a pairwise distinguishability attack against $B$ has advantage bounded by $(e^\varepsilon-1)/4 + \delta/2$.
\end{restatable}

The proof can be found in Appendix \ref{appendix:calculations}. For an example, consider $\varepsilon=0.5,\delta=0.01$: advantage is bounded by $<0.17$ which normalizes to $<0.34$.

Our definition of distinguishability attack can be compared to membership inference attacks~\cite{homer2008, sankararaman2009, dwork2015robust}. In both cases, an attacker wants to learn a binary predicate about the target. The predicate is membership for membership inference, while our predicate concerns value.

\subsection{Fingerprinting Attacks}
\label{sec:threat_model_fingerprinting}
One way to generalize pairwise distinguishability is $k$-wise distinguishability, where prior knowledge $W$ covers a large set $\{x_1, x_2, \dots, x_k\}$. We additionally refer to an \emph{interest set} $I$. To continue our URL example, the adversary may be interested in URLs that end in country codes.
The adversary has two objectives: to determine whether the target's URL is interesting and, if it is, to identify which interesting URL it is. The country code can serve as a hint about the target's location or language. Reconstruction of uninteresting URLs is not a priority.

Similar to the pairwise distinguishability attack, the prior $W$ grants the adversary baseline strategies that do not involve side channels at all. Specifically, to determine whether the target $x$ is in $I$, the baseline strategy is to report ``interesting'' if and only if the mass placed on set $I$ by $W$ is larger than the mass placed outside it. In our example, this amounts to comparing the prior probability of visiting a URL with a country code against that of visiting a URL without one. To fingerprint $x$ assuming it is in $I$, the baseline strategy is to report the likeliest element according to the distribution $W_I$, which is $W$ conditioned on $I$; this amounts to reporting the most frequently visited URL ending in a country code. Note that the success rate of the baseline interesting/not-interesting classifier is $s_c := \max(\pr{x \leftarrow W}{x\in I}, \pr{x \leftarrow W}{x\notin I})$, while the baseline fingerprinting success rate is $s_f := \max_{i \in I} \pr{x \leftarrow W_I}{x=i}$.

With side-channel leakage, we again would like to bound the advantage over these baseline rates. The adversary observes side channels via the fingerprinting game:

\begin{definition}[Fingerprinting Game]
Assuming prior $W$, the fingerprinting game proceeds as follows:
\begin{compactenum}
    \item (Input Preparation) Target user's data $x$ randomly chosen according to $W$
    \item (Before Input Ingestion) Attacker chooses Sybils $X$ and chooses $I$
    \item (Input Ingestion) Whole dataset $D $ is formed by appending $x$ to $X$
    \item (After Report) $\attackeroutput_\attacker := \attacker(\sidechannels_\attacker^B(D), W, I)$
\end{compactenum}
\label{def:finger-game}
\end{definition}

The interest-classification advantage is $\advantage_\attacker := \max(0, \pr{x\leftarrow W}{(\attackeroutput_\attacker == \textrm{``interesting''}) = x\in I} - s_c)$. We can normalize this advantage to the range $[0,1]$ by dividing by its maximum value $1-s_c$. The fingerprinting advantage is $\max(0, \pr{x \leftarrow W_I}{\attackeroutput_\attacker = x} - s_f)$. We can again normalize by dividing by its maximum value $1-s_f$.

\begin{example}[Fingerprinting Game]
Suppose an attacker picks $ I = \{ \text{any example site} \neq \text{'example.com'}\}$
and the prior probability distribution for visiting a website is

    \begin{tabular}{c|c}
         60\% example.com & 10\% example.co.jp\\ \hline
         10\% example.co.uk & 20\% other URLs\\
    \end{tabular}

The real site visited ($x$) is drawn randomly from $W$. The baseline chance of guessing (non-)membership in $I$ is $s_c = \max(0.2,0.8)=0.8$; absent side-channels, the best guess is that the target is not in $I$. The baseline chance of reconstructing an element of $I$ is $0.5$; absent side-channels, there is an even chance between \texttt{example.co.uk} and \texttt{example.co.jp}.

If the side-channel leakage observed by the attacker ($\attackeroutput_\attacker$) leads to a $0.9$ probability of guessing whether $x\in I$, then the interest-classification advantage is $0.1=0.9-0.8$. If the leakage grants a $0.7$ chance of recovering an element of $I$, the attacker has a fingerprinting advantage of $0.2=0.7-0.5$.
\end{example}

In our empirical evaluation of TF-Federated's PHH implementation (Section \ref{sec:phh}), we set $W$ to be the actual distribution of the
data going into the CVM, thus assuming the adversary has perfect prior knowledge about the distribution. Note that this sets a high bar for what 
constitutes a successful side-channel attack
to reconstruct a victim's input.

%% file: 04.framework.tex
\section{\FrameWorkName\ Framework} 
\label{sec:framework}

\begin{figure}[t!]
    \centering
        \includegraphics[width=\linewidth]{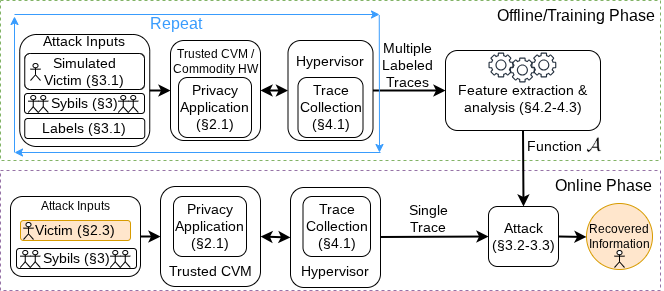}
    \caption{The overview of \FrameWorkName. The corresponding section numbers $\S$ are listed for each component. The offline phase consists of building an ML attacker model using labeled traces (Section \ref{sec:leakage-analysis}). The online phase uses this model on a single CVM trace that includes victim data, to launch a privacy attack $\mathcal{A}$ that reveals this victim data.}
    \label{fig:system_diagram}
\end{figure}

Figure~\ref{fig:system_diagram} presents an overview of \FrameWorkName. The offline phase allows developers to model attacks of various strengths and capabilities, and analyze privacy leakage under these instantiations. This is achieved by collecting a labeled dataset of traces corresponding to the attack, and using statistics and ML tools to build leakage-analysis tools. In the online phase, the leakage-analysis tool is used to conduct an attack $\mathcal{A}$ on a single trace containing the victim data, to quantify privacy leakage. This resembles an attacker who collects side-channel information offline to build a leakage detector for online use, on a victim CVM through a malicious hypervisor. We present how \FrameWorkName implements the collection of different side-channel signals, feature extraction, and the leakage-analysis tools. 

\subsection{Trace Collection}
\begin{figure}[t!]
    \centering
    \includegraphics[width=0.9\linewidth]{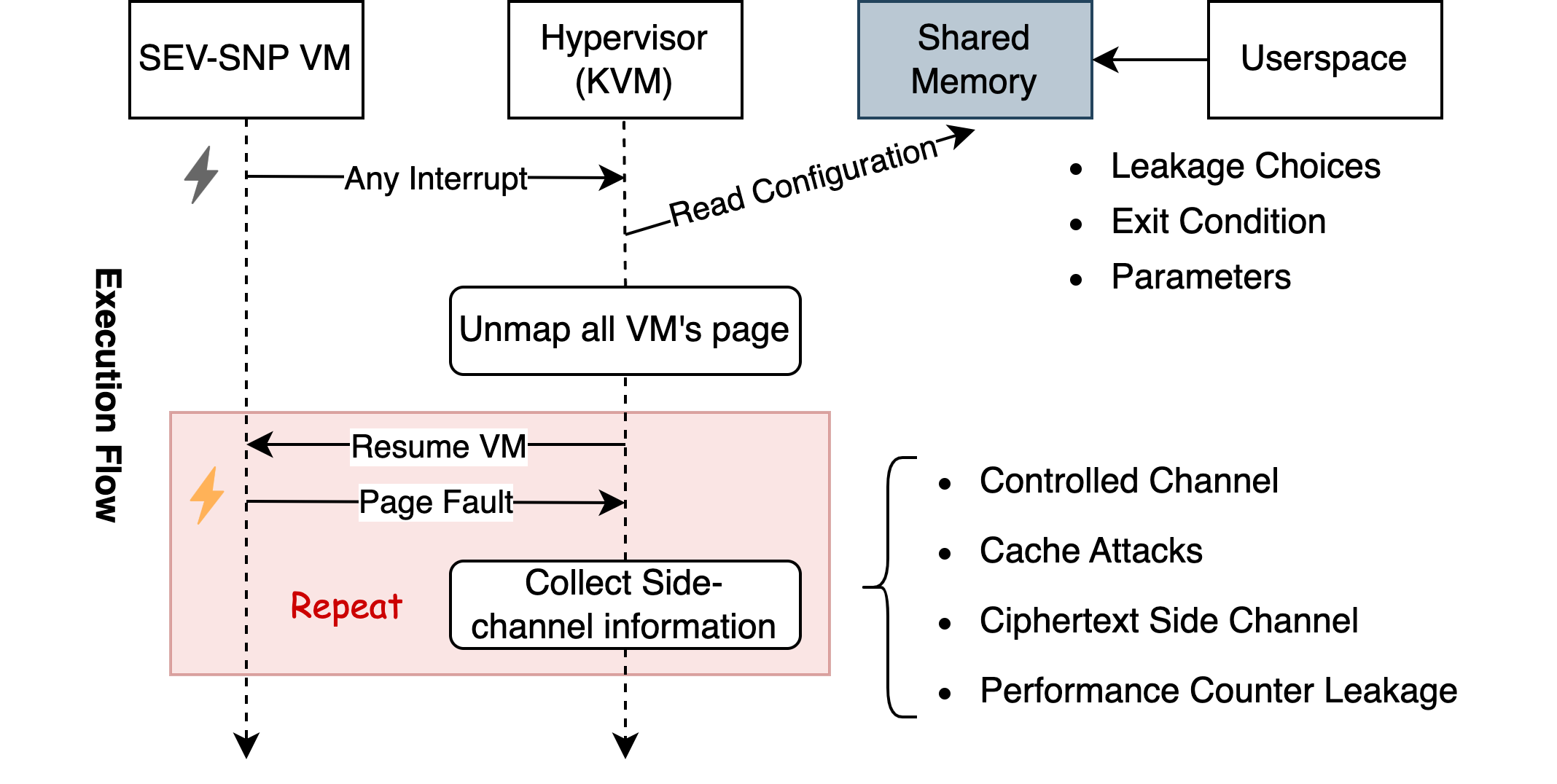}
    \caption{The overview of \FrameWorkName\ trace collection. A user-space controller uses shared memory to configure each collection run with a modified KVM module. Grey lighting marks any interrupt allowing the hypervisor to read the configuration; yellow lighting is a page fault interrupt as the hypervisor clears the present bit of guest pages.
    }
    \label{fig:overview}
\end{figure}

\Cref{fig:overview} gives an overview of the side-channel trace collection in \FrameWorkName, which consists of a modified KVM module and a user-space controller, communicating through shared memory. Besides configuring trace collection runs (e.g., via number of traces, optimizations), the shared memory enables a developer to control the two key components of trace collection: \textit{temporal resolution} -- determining the frequency of side-channel event collection, and \textit{spatial resolution} -- the side-channels with different granularities. 

Nevertheless, in contrast to cryptographic targets, there are significant challenges in automating trace collection and analysis in privacy applications. This requires strategies to improve collection speed by ignoring uninteresting parts of the execution flow, mapping only relevant code pages, and supporting generic analysis through various attack primitives. \FrameWorkName achieves this through two \textit{key insights} during temporal and spatial resolution, which allow developing general and platform-specific optimization strategies.

\subsubsection{Temporal Resolution}
We introduce our technique to synchronize the hypervisor's side-channel collection with the target running inside the SEV-SNP VM. 
The hypervisor requires a trigger point to halt the VM execution and control its execution.
Well-known methods include inducing page faults~\cite{xu2015controlled}, triggering interrupts with APIC timers~\cite{van2017sgx,wilke2023sev}, or using a combination of both~\cite{moghimi2020copycat,zhang2024cachewarp}.
APIC interrupts allow the attacker to pause the VM at short intervals, ensuring that only one instruction is completed after each context switch, known as single-stepping~\cite{van2017sgx}.

While single-stepping might seem like an obvious solution for temporal resolution, it is suboptimal for privacy attacks for two reasons. 
First, single-stepping is inherently slow. 
Each step requires at least one context switch, and it takes even longer when zero stepping occurs (i.e., no progress after a context switch). 
For example, our fingerprinting testcase (see Section~\ref{sec:phh-advanced}) completes in \SI{0.46}{\milli\second} without monitoring.
With a filtered controlled channel, \FrameWorkName collects \num{1.06e5} entries in \SI{0.56}{\second}, incurring approximately \num{1217}x overhead, which remains acceptable for offline auditing.
In contrast, after integrating single-stepping, \FrameWorkName generates \num{2.8e5} zero-step, \num{1.6e6} single-step, and \num{6.7e5} two-instruction entries in \SI{8.67}{\second}, yielding an approximately \num{18848}x overhead.
Such high overhead typically necessitates narrowing analysis to a limited code section, as in cryptographic libraries~\cite{li2022systematic,teeJam2023,moritz2021platypus,Puddu2021frontal}.
However, privacy-preserving applications often have code bases much larger than cryptographic libraries, making such an analysis prohibitively time-consuming.
Importantly, unlike in cryptographic libraries -- where the attacker's goal can be clearly identified (i.e., stealing the secret keys), privacy-preserving applications generally do not reveal where leakage may occur, making binary analysis much more challenging. 

Second, single-stepping could be restricted in future architectures. While current tools can use performance counters to detect if a single-step was successful~\cite{wilke2023sev}, AMD claims to prevent the hypervisor from reading performance counters for guest events starting with Zen 5~\cite{amczen5hpc}. 
Moreover, Intel TDX mitigates single-stepping by ensuring sufficient VM progress between context switches~\cite{constable2023aex}. 
Therefore, although supporting single-stepping is not a limitation of our framework, it was not included in our evaluation.

\underline{\textbf{Key Insight 1: Filtered Controlled-Channel.}} We choose nested page faults for \FrameWorkName, which, besides being much faster, represent a design choice that cannot be mitigated without major architectural changes. 
As shown in \Cref{fig:overview}, the untrusted hypervisor clears the present bit of all the VM pages at the beginning of the execution.
When the guest VM triggers page faults, the error code reveals whether the faulting page is used for instruction or data and whether it is encrypted.
If the page is not encrypted, the hypervisor considers it uninteresting and retains its mapping in subsequent executions.
Similarly, if a faulted page address belongs to the reserved memory of the guest system, it is likely associated with kernel activity and can be skipped.
To ensure accurate control and data flow tracking, \FrameWorkName maps the newly accessed page and conditionally unmaps the previous one at each page fault.

\textbf{Platform-specific optimizations.} To further refine page-level analysis, \FrameWorkName also supports additional platform-specific optimizations. For AMD EPYC CPUs, we leverage performance counters that allow the hypervisor to monitor guest events in either user-space or OS-space. Our implementation uses two of these: one tracking retired instructions from guest user-space and another tracking retired micro-operations (uops) from the guest OS. As a result of these optimizations, pages containing only kernel code are labeled irrelevant and excluded from analysis.
This optimization is optional to improve runtime on AMD EPYC, and is not necessary for page-level analysis. As a result, \FrameWorkName also generalizes to other architectures~\cite{aktas2023intel}. We analyze the collection speed in \Cref{table:col-speed} (Appendix~\ref{appendix:framework_and_ml}).

\subsubsection{Spatial Resolution}
We use \FrameWorkName\ to collect runtime side channels of the target program at granularity levels ranging from 4\,kB to 16\,B. These include both generalizable platform-independent ones (controlled-channel and cache), and platform-dependent ones (ciphertext and PMCs). \Cref{fig:trace} in Appendix~\ref{appendix:framework_and_ml} shows an example trace with collected side channels.

\para{Platform-independent: Controlled-channel.}
We monitor access patterns of the victim at page granularity with controlled-channel techniques, distinguishing between code fetches and data accesses.
\FrameWorkName\ follows a principle of mapping only one \textit{interesting} code page at a time, unmapping the current code page whenever the guest jumps to a new one\footnote{Corner cases, such as a single instruction spanning a page boundary, can be handled by detecting a threshold of two repeated page faults without execution progress, using PMCs.}.
For data accesses, \FrameWorkName\ manages a queue for mapping data pages.
The queue size is adjusted dynamically to avoid deadlocks when a single instruction accesses more pages than the queue size.
This ensures precise control over memory access patterns, without losing track of any accessed pages.

\para{Platform-independent: Cache attacks.}
While controlled-channel techniques have a 4\,kB page granularity, access patterns with a 64\,B granularity are possible via cache attacks.
When handling an NPF, the hypervisor iterates the nested page table and maps the faulted page before returning control to the VM.
As an attacker cannot predict which 64\,B blocks of a page the victim accesses, we mount a Multi-Prime+Probe on all 64 cache sets before the context switch.
At the next NPF, the hypervisor probes all cache sets to identify the accessed 64\,B blocks. 

Developing a precise Multi-Prime+Probe attack poses significant challenges.
On newer AMD CPUs, such as EPYC, the shared L3 last-level cache is non-inclusive. Therefore, performing L3 Prime+Probe on each cache set requires accessing at least 24 addresses, accounting for both L2 and L3 cache set entries~\cite{Zhang2025Allocator}.
Although the untrusted hypervisor shares an internal L2 cache with the target VM, the timing difference between L2 hits and misses is only about 40 cycles.
This small difference introduces significant noise when attempting to probe across 64 different cache sets, as we will see below. 

\underline{\textbf{Key Insight 2: Noise reduction via MSRs.}} We introduce a novel approach to improve L2 Prime+Probe by exploiting model-specific-registers (MSRs) to reserve L3 cache usage.
The MSRs \texttt{0xC001\_1095} and \texttt{0xC001\_1096} define a memory range for which the number of L3 ways can be configured via the MSR \texttt{0xC001\_109A}.
We allocate our buffer for eviction set at a high memory address beyond 30\,GB, and use the MSRs to reserve all L3 ways for memory addresses below 30\,GB.
Consequently, the addresses from our eviction set cannot be cached in the last-level cache, as there is no available way.
Hence, if an address in the eviction set is evicted from the L2 cache, it is directly evicted to the main memory, increasing the measured timing difference for Multi-Prime+Probe to more than 700 cycles.
This approach improves the speed of the prime stage from 24 distinct addresses accessed to only 8 (the capacity of an L2 cache set), without introducing noticeable performance overhead, as the memory space above 30\,GB is rarely used.

While these specific MSRs were introduced in the Zen3 microarchitecture, similar cache reservation technologies (e.g., Intel's Cache Allocation Technology, or CAT) can achieve a comparable effect on older AMD CPUs and Intel CPUs. 
We demonstrate a proof-of-concept improving Prime+Probe using Intel CAT in Appendix~\ref{appendix:intelcat}.

\para{Platform-specific: Ciphertext.}
Similar to cache attacks, before the guest writes to an unmapped page, the hypervisor reads all 256 ciphertext blocks of the page~\cite{li2021cipherleaks}. 
At the next NPF, \FrameWorkName compares all blocks to their prior values to pinpoint the ones modified by the victim, and highlight ciphertext differences. This exploit provides insights into the victim code at an even deeper spatial side-channel level.

\para{Platform-specific: PMC leakages.}
In addition to the two events we introduce for optimizing temporal resolution on AMD EPYC, \FrameWorkName uses three other events used by Gast et al.~\cite {gast2025counterseveillance}, \textit{Retired Branch Instructions}, \textit{Retired Taken Branch Instructions}, \textit{Retired Near Returns}, that leak the control flow of the victim. 
The former is updated at each NPF, and the latter during instruction fetch NPFs. 
This method leverages performance counters to provide detailed insights into the execution flow, further enriching the side-channel data available for analysis.

\subsubsection{Targeted Trace Collection}\label{sec:targeted-trace-collection}

In addition to tracking across the entire target call flow, \FrameWorkName enables precise leakage analysis by offering configurable controls over the monitoring phase.
This allows developers to instrument applications with code pages containing specific assembly instructions, provided these instructions can be tracked by performance counter events. 
In the offline phase, the developer wraps the target code snippet with two code pages that execute the \texttt{clflush}\xspace instruction multiple times, enabling the framework to treat these pages as signals to start or stop tracking.
As we show in \Cref{sec:udf}, in real-world exploits, this is equivalent to an attacker who would monitor specific I/O, network traffic, and access patterns in the online phase.

\subsection{Feature Extraction}

\begin{table*}[ht]
\begin{center}
\resizebox{0.86\linewidth}{!}{
\begin{tabular}{ llll}
\toprule
 \textbf{Name} & \textbf{Feature Set} & \textbf{Description} & \textbf{Count} \\ \midrule
 CF Count & $\mathcal{F}_{1}$ & Number of total \& unique code pages fetched & 2 \\  
 DA Count & $\mathcal{F}_{1}$ & Number of total \& unique data pages accessed & 2\\  

Cache Count & $\mathcal{F}_{2}$ & Number of total \& unique intercepted cache lines in cache attacks & 2\\  
 CI Count & $\mathcal{F}_{2}$ & Number of total \& unique modified ciphertext blocks in memory & 2\\  
 
Cache Frequency & $\mathcal{F}_{3}$ & Number of times each of the 64 4\,kB cache lines was accessed & 64\\ 
 CI Frequency & $\mathcal{F}_{3}$ & Number of times each of the 256 blocks of any page was modified & 256\\ 
 
 DA Stats & $\mathcal{F}_{4}$ & Stats over the number of data page accesses following a code page fetch & 11+N \\ 
 Cache Stats & $\mathcal{F}_{4}$ & Stats over the number of total \& unique cache lines accessed for a page & 2*(11+N) \\ 
 CI Stats & $\mathcal{F}_{4}$ & Stats over the number of total \& unique blocks accessed for a page & 2*(11+N) \\ 
 
 CF Page Frequency & $\mathcal{F}_{5}$ & Frequency of code fetches for individual code pages & M$_{CF}$ \\  
 DA Page Frequency & $\mathcal{F}_{5}$ & Frequency of page accesses for individual data pages & M$_{DA}$ \\
 \bottomrule
\end{tabular}}
\caption{Summary of handcrafted features in \FrameWorkName. $\mathcal{F}_{4}$ Stats correspond to features describing the distribution: min, max, first to ninth quantiles, and a histogram with N bins. $\mathcal{F}_{5}$ frequencies are over the M first seen pages in a trace. }
\label{table:handcrafted_features}
\end{center}
\end{table*}

Given a set of execution traces, \FrameWorkName\ extracts features to model privacy attacks as discriminative machine learning problems, which aim to separate traces depending on the value of the target.
The types and complexity of features chosen often result in a trade-off between interpretability and utility. 
Simpler features tend to be more suitable for pinpointing the source of the leakage, while high-level features are capable of producing a tighter empirical lower bound of the side channel. 
To support the automated audit process and balance this trade-off, \FrameWorkName\ employs both handcrafted features and automatic feature learning. 

\subsubsection{Handcrafted features} 
We engineer five sets of handcrafted features as summarized in Table~\ref{table:handcrafted_features}. 
These features do not exhaustively capture all the distinguishing patterns that can be extracted from traces, but, as we show in Sections~\ref{sec:pir} and~\ref{sec:phh}, highlight the utility of the framework in pinpointing sources of leakage.
Feature set $\mathcal{F}_{1}$ focuses on a page-level granularity and counts the number of total and distinct pages observed for the controlled channel, across a particular trace, while $\mathcal{F}_{2}$ operates at block- and cache-level granularity. $\mathcal{F}_{3}$ captures a lower level of spatial granularity by computing histograms for the number of times individual cache lines and page blocks are accessed in cache and ciphertext side channels, respectively. $\mathcal{F}_{4}$ looks at the side channel for each individual page level, computing how many data accesses, cache lines, and blocks are being accessed, and providing statistics for these across a trace. $\mathcal{F}_{5}$ captures aspects of the control and data flow by counting how many times individual pages are accessed during the execution. 

\subsubsection{Automatic feature learning} 
Alternatively, we rely on representation learning, allowing analysts to train deep learning models for automated analysis of the leakage. %
This involves feeding the trace information to the models and relying on their representational power to expose discriminative features from the sequences encoded in the traces.
To aid learning, we pre-process the traces by abstracting the memory space and ciphertext information. More precisely, we extract only the distinct page-table accesses (code and memory pages) and the ciphertext changes observed through the ciphertext visibility channel. This transformation is described in Appendix~\ref{appendix:framework_and_ml}.

\subsection{Leakage Analysis}
\label{sec:leakage-analysis}
To conduct a leakage discovery task, one can build different analysis models using the framework. We implement several analytics and machine learning tools for evaluating features against datasets and identifying leakage.
Collected traces first need to be separated into different classes to define a classification problem. 
The labeling depends on the threat model. 
A distinguishing attack (Section~\ref{sec:threat_model_distinguishability}) can be modeled using two classes, indicating whether the targeted entity is present in the input set for a particular trace. 
In contrast, fingerprinting attacks (Section~\ref{sec:threat_model_fingerprinting}) can be modeled as two consecutive distinguishing problems: 
a two-class setting reflecting whether the targeted entity is part of a set of labels of interest, followed by a multi-class setting that identifies which of the labels the entity corresponds to.

After collecting and labeling the traces, the frameworks can discover the source and severity of the leakage across different labels. This is achieved by choosing the feature sets, localizing the portion of the program suspected of leakage, and computing features over the collected traces. After feature extraction, the leakage-analysis tools are applied. To identify leakage sources, we provide statistical tests and visualization tools to verify whether the distributions of features across labels are distinct. For quantifying the leakage, we implement supervised learning models that allow measuring leakage in the online phase. Our implementation relies on scikit-learn~\cite{scikit-learn} for classifiers based on handcrafted features, and on TensorFlow~\cite{tensorflow2015-whitepaper} for feature learning through sequence models.

\para{Evaluation} To measure the leakage and highlight the empirical advantage through ~\FrameWorkName, in Sections~\ref{sec:pir} and~\ref{sec:phh} we use an L2-regularized logistic regression classifier trained for 1000 iterations with L-BFGS. For feature learning, we implement a bidirectional LSTM with attention (see Appendix~\ref{appendix:framework_and_ml}). 
The model has \num{751554} parameters and is trained with Adam using a learning rate of 2e-5. We compute the empirical advantage by training and validating on 80\% of the samples and testing on the remaining 20\%. For the logistic regression, we report the average over 5 trials. 

%% file: 06.pir.tex
\section{Oak Private Information Retrieval} \label{sec:pir}
In this section, we evaluate \FrameWorkName on \emph{Oak}~\cite{oak,oakSLSAblog} private information retrieval (PIR).
Project Oak is a software platform developed by Google for constructing distributed systems with built-in transparency and guarantees of confidentiality and integrity.
It provides core components for developing enclave applications and supports remote attestation.
Oak includes an untrusted launcher on the host and uses a Wasm runtime to execute Wasm enclave applications within a CVM.
The launcher handles requests using gRPC, providing end-to-end encryption for data in transit.
This architecture supports PIR for in-memory key-value lookups, enabling sensitive data queries while preserving the confidentiality of data and queries.

PIR generally allows a client to retrieve an element from a (typically public) database, without revealing the accessed element to the server that hosts the database.
Cryptographic solutions to PIR are well-established~\cite{chor1998private}, but even the most efficient constructions fundamentally require the server to scan the entire database to answer a query, which limits the scalability.
To overcome these issues, many solutions~\cite{oak,signal-blog,mokhtar2017x} instead use a TEE with the goal of protecting the queried index from the server.

\subsection{Distinguishing Attack on Oak PIR}

\begin{figure}[tb]
    \centering
    \includegraphics[width=\columnwidth]{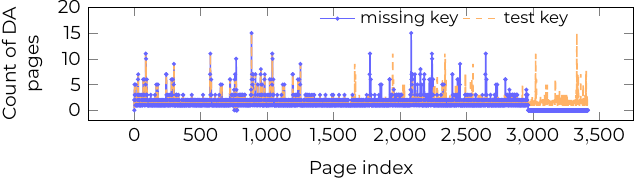}
    \caption{Traces of the Wasm runtime for the oak \texttt{key\_lookup} module with a missing and a test key~\cite{oakLookup}.}
    \label{fig:oak-lookup}
\end{figure}

As a motivating example, we distinguish the request of a missing key and a key with a value, capturing the resulting traces on the hypervisor. 
\Cref{fig:oak-lookup} shows sequences of the number of data page accesses at each code page. The traces are clearly distinguishable, and we omit a more detailed analysis using machine learning.

For a more realistic scenario, we expand the dataset to include 1,000 key pairs. The keys are the strings \texttt{key0} through \texttt{key999}, resulting in lengths of 4 to 6 bytes. Each corresponding value is randomly generated with a size of up to 1,000 bytes.
Although the sequences shown in \Cref{fig:oak-lookup} remain the same across these keys, we can still distinguish individual lookups by observing variations in page and cache accesses at specific code pages, as shown in \Cref{table:oak-key1000}.
The lookup module must load the key-value pairs from different pages containing distinct cache lines for the second to fifth data page accesses from this code page. 
The number of cache-line accesses also reflects the size of the values, which can be up to 1,000 bytes and span multiple cache lines. 
Note that we use manual examination to showcase and validate this leakage. 
We collected the traces using Targeted Trace Collection~(\Cref{sec:targeted-trace-collection}). 
Fully automating the end-to-end exploit would require an attacker to recognize a pattern targeting the vulnerable code and train a new model offline.

Following Definition~\ref{def:dist-game}, the goal for the attacker is to distinguish two sequences of memory accesses into a database of 1000 elements: one consisting of 10 identical PIR retrievals of the first element, and the other consisting of 9 identical retrievals of the first element, followed by a single retrieval of the last element.
We collect 1500 traces for each case and use them to evaluate the effectiveness of using \FrameWorkName to evaluate PIR mitigations.

\begin{table}[t]
\begin{center}
\resizebox{0.9\linewidth}{!}{
\begin{tabular}{ c|cccccc}
\toprule
 & \textbf{key0} & \textbf{key1} & \textbf{key60} & \textbf{key61} & \textbf{key998} & \textbf{key999}  \\ \midrule
1st & 123dce & 123dce & 123dce & 123dce & 123dce & 123dce \\  
    & \textit{31} & \textit{31} & \textit{31} & \textit{31} & \textit{31} & \textit{31} \\  
2nd & 106835 & \color{blue}{106aea} & \color{green!40!black}{106ae9} & \color{green!40!black}{106ae9} & 106835 & 106835 \\
    &  &  &  &  &  &  \\  
3rd & 106b9e & \color{green!40!black}{123e6e} & \color{green!40!black}{123e6e} & \color{green!40!black}{123e6e} & \color{blue}{106b9d} & \color{blue}{106b9d} \\  
    &  &  &  &  &  &  \\  
4th & \color{green!40!black}{123e6e} & 10696c & \color{blue}{10696b} & \color{blue}{10696b} & \color{green!40!black}{123e6e} & \color{green!40!black}{123e6e} \\
    &  &  &  &  &  &  \\  
 \textbf{5th} & \textcolor{black}{\textbf{139213}} & \textcolor{green!40!black}{\textbf{142816}} & \textcolor{blue}{\textbf{13a018}} & \textcolor{red!80!black}{\textbf{139010}} & \textcolor{cyan}{\textbf{136617}} & \textcolor{orange!80!black}{\textbf{136a1a}} \\  
    & \textcolor{black}{\textit{\textbf{48,49,63}}} & \textcolor{green!40!black}{\textit{\textbf{7,16-31}}} & \textcolor{blue}{\textit{\textbf{32,50-52}}} & \textcolor{red!80!black}{\textit{\textbf{58,60,63}}} & \textcolor{cyan}{\textbf{0-6,22,62}} & \textcolor{orange!80!black}{\textbf{6,32}} \\  
6th & 123dce & 123dce & 123dce & 123dce & 123dce & 123dce \\  
    & \textit{31} & \textit{31} & \textit{31} & \textit{31} & \textit{31} & \textit{31} \\  
\end{tabular}}
\caption{An example of distinct page- and \textit{cache-access} sequences appears in one of the code pages (index 1,642) within the oak \texttt{key\_lookup} trace for six different key lookups. The page number refers to the guest's physical page number (gPN), followed by cache-line accesses within this page. We repeat the lookup on each key five times.}
\label{table:oak-key1000}
\end{center}
\end{table}

\subsection{Evaluating Mitigations}
To mitigate the above leakage, we can apply a linear scan, i.e., a \texttt{std::vector} accessed through a scan, using a constant-time compare-and-swap, and the PathORAM~\cite{stefanov2018path} used by Signal~\cite{signal-blog}. 
We run both applications inside a CVM. 
We collect traces with \FrameWorkName and evaluate the collected traces using the methods described in Section~\ref{sec:framework}.

\para{Distinguishing attack via handcrafted features.}
First, we analyze the leakage exposed via handcrafted feature sets summarized in Table~\ref{table:handcrafted_features}. 
We train a logistic regression model on each of the feature sets, as well as their union.
The results are summarized in Table~\ref{table:pir_results}.
While none of these features reveals a significant advantage against the constant-time linear scan implementation, feature sets $\mathcal F_5$ reveal leakage in Signal's ORAM. 
The leakage originates from a ciphertext side-channel related to how zero-value items are handled.
When an item with a zero value is retrieved, it is added back to memory, leaving the ciphertext unchanged. 
In contrast, retrieving a non-zero value causes the plaintext to change, which results in a changed ciphertext.
This is consistent with the observation that retrieving different non-zero values produces no discernible difference in the trace.


\para{Distinguishing attack via feature learning.}
To explore what information can be observed without any feature engineering, we also train an LSTM model on the pre-processed traces.
The advantages obtained through the LSTM on the test dataset are shown in Table~\ref{table:pir_results}.
Using only page-level information, the LSTM cannot get any
significant advantage against the two mitigations.
However, once we add the ciphertext block-level visibility side channel, we observe that both linear scan and ORAM are distinguishable. 
This finding aligns with the handcrafted features-based analysis, reinforcing the vulnerability in these implementations: side-channel visibility into the exact ciphertext changes reveals more than just the number of changes. Nevertheless, the larger attacker advantage obtained through automatic feature learning over handcrafted features highlights the complementary role of the two in \FrameWorkName: while handcrafted features are useful for pinpointing the source of the leakage, feature learning provides tighter empirical advantage estimates. 

\para{Remark.} While our experiments highlight leakage through the ciphertext channel, this is due to our experiments using AMD SEV-SNP, which is known to be vulnerable. In contrast, Signal's ORAM was implemented with Intel SGX as its target architecture, which does not suffer from this side channel in the same way. We therefore cannot confirm any vulnerability in Signal's deployment of ORAM.

\begin{table}[t]
\begin{center}
\resizebox{\linewidth}{!}{
\begin{tabular}{ l|llllll|lc}
\toprule
 & \multicolumn{6}{c|}{Logistic Regression} & \multicolumn{2}{c}{LSTM} \\ 
 & \textbf{$\cup$} & \textbf{$\mathcal{F}_{1}$} & \textbf{$\mathcal{F}_{2}$} & \textbf{$\mathcal{F}_{3}$} & \textbf{$\mathcal{F}_{4}$} & \textbf{$\mathcal{F}_{5}$} & \small{Page} & \small{Page+Block}  \\ \midrule
Linear Scan & \textbf{0.02} & 0.00 & 0.00 & 0.01 & 0.01 & \textbf{0.02} & 0.03 & \textbf{0.80} \\  
Signal ORAM & \textbf{0.19} & 0.03 & 0.04 & \textbf{0.18} & 0.02 & 0.07 & 0.03 & \textbf{0.32} \\
 \bottomrule
\end{tabular}}
\caption{Normalized advantage (Definition~\ref{def:dist-game}) obtained through a Logistic Regression on all ($\cup$) and sets of ($\mathcal{F}$) handcrafted features, and an LSTM on sequence-based features at page- and block-level, across PIR implementations. }
\label{table:pir_results}
\end{center}
\end{table}

%% file: 07.phh.tex
\section{Private Heavy Hitters in TF-Federated} 
\label{sec:phh}

The private heavy hitters (PHH) problem has received immense research attention, with cryptographic solutions deployed under multiple threat models.
PHH aims to compute a histogram of the user data, while providing a (differential) privacy guarantee to individual users, with $n$ users $1, \ldots, n$ each holding one datapoint from some large domain. 
Recent deployments by Google~\cite{tfblogpost} and Meta~\cite{srinivas2024federated} leverage AMD SEV-SNP and Intel SGX for this task, respectively.

In this section, we evaluate \FrameWorkName on the TensorFlow-Federated~\cite{fedTFphh} implementation of Private Heavy Hitters in Confidential VMs by Google, as deployed in Gboard via AMD SEV-SNP~\cite{tfblogpost}. The specific application is out-of-vocabulary word discovery: ``discovering new common words to incorporate them into the typing model, without revealing any uncommon private words.'' 
The work of Srinivas et al.~\cite{srinivas2024federated} describes a deployment of the same algorithm based on SGX, but does not discuss mitigations to architectural side channels nor make any code available. 

\para{Algorithm for DP Heavy Hitters.} 
TEE-based solutions for PHH~\cite{eichner2024confidential,tfblogpost,srinivas2024federated} apply a textbook DP algorithm inside a TEE~\cite{salilbook} (the so-called stability-based histograms or noise-and-threshold), and rely on the TEE to safeguard inputs and keep sampled DP noise confidential. 
In particular, the TensorFlow-Federated implementation~\cite{fedTFphh, tfblogpost} closely follows the textbook DP mechanism for large domain histograms. Figure~\ref{fig:vanilla-phh} shows a basic C++ reference
implementation, for illustrative purposes.

\begin{figure}[t]
\begin{minted}{c}
// batch, epsilon, and threshold in the context
std::unordered_map<string, int> hist;
std::vector<std::pair<string, int>> result;
...
while(!batch.empty()){ // Aggregate inputs
  hist[batch.front()]++;
  batch.pop();
}
for (auto [k, v]: hist) { // Noise and Threshold
  auto noisy_val = v + sample_centered_laplace(epsilon);
  if (noisy_val >= threshold)
    result.push_back(std::make_pair(k, noisy_val));
}
\end{minted}
\caption{Baseline unprotected PHH application example code. ``\texttt{batch}'' refers to user data to be protected. The variance of the noise and the threshold are set according to $\epsilon$ and $\delta$, to achieve $(\epsilon, \delta)$-DP.}\label{fig:vanilla-phh}
\end{figure}

\para{Case Study: Counting Common URLs.}
A typical application of PHH is in browser telemetry~\cite{poplar, mastic}, where clients report URLs that crashed their browser and the related context (see Network Error Logging~\cite{network-error-logging}). 
In the rest of this section, for illustrative purposes, we consider a DP algorithm that attempts to identify frequent URLs submitted by devices. The goal of the attacker is to extract additional information (beyond the result histogram) about the URLs submitted by individual devices, as formalized in Definitions~\ref{def:dist-game} and~\ref{def:finger-game}. This attack model also incorporates Sybil attacks (\Cref{sec:threat_model}).

\begin{figure*}
\begin{subfigure}[t]{0.49\hsize}
\includegraphics[width=\textwidth]{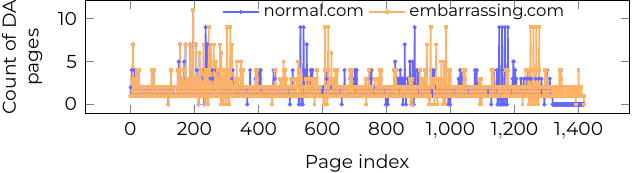}
\caption{Last iteration of the accumulation phase}\label{fig:tf-phh-acummulate}
\end{subfigure}
~
\begin{subfigure}[t]{0.49\hsize}
\includegraphics[width=\textwidth]{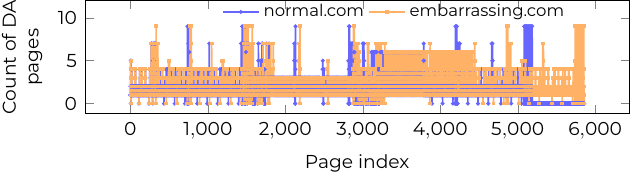}
\caption{The noise-and-threshold phase}\label{fig:tf-phh-report}
\end{subfigure}

\caption{Page-level leakage of TensorFlow Federated PHH code during accumulation and noise-and-threshold phases. We ingest ``normal.com'' $9$ times, then either ``normal.com'' or ``embarrassing.com''. Adding a new key triggers a longer code path in accumulation and an extra iteration in the noise-and-threshold phase.}
\end{figure*}

\subsection{TF-Federated Evaluation}
We analyze the leakage of the aggregation and noise-and-threshold phases of the DP algorithm\footnote{Source code available at: google-parfait/tensorflow-federated, file: dp\_open\_domain\_histogram\_test.cc, line 655, commit e245ed4}.
As in the snippet in Figure~\ref{fig:vanilla-phh}, the aggregation phase accumulates inputs into a hash map and places the keys into a vector.
The noise-and-threshold phase iterates through the vector, adding DP noise to each histogram entry, and then thresholding.

\para{Findings.} In both parts of the code (aggregation and noise-and-threshold), the code leaks sufficient information 
to enable a distinguishing attack (Definition~\ref{def:dist-game}).
Figures~\ref{fig:tf-phh-acummulate} and~\ref{fig:tf-phh-report} show page-level leakage in the application.
We plot the count of data page accesses after each code page fetch (feature set $\mathcal{F}_4$) in two neighboring executions: in the first one, we ingest ``normal.com'' $10$ times, while in the second one, we ingest into the application 
``normal.com'' $9$ times, followed by ``embarrassing.com''.
This corresponds to an instance of a distinguishing attack from Definition~\ref{def:dist-game}. 
The plots show that the execution length is 
data-dependent for both the aggregation and noise-and-threshold phases. We examine this further, along with a potential mitigation, in the following section.

\subsection{Advanced Attacks and Mitigations}\label{sec:phh-advanced}
In the rest of this section, we discuss different flavors of attacks and mitigations, and how their effectiveness can be evaluated with \FrameWorkName. Instead of working with the TF-Federated codebase from the previous section, the results in this section are with respect to a smaller example (partially reported in Figure~\ref{fig:vanilla-phh}) and included with our library. This simpler example is less leaky than a larger codebase, and allows us to better identify the origin of the leakage when using automatic ML approaches to exploit it. Moreover, any attack found in the simpler codebase translates to the more complex TF-Federated implementation, and mitigations are easier to implement and evaluate.

We start by discussing an attack based on data-dependent execution. This attack is analogous to the one reported in Figure~\ref{fig:tf-phh-report}, and mentioned above.

\begin{figure}[t]
\centering
\begin{subfigure}[t]{.49\columnwidth}
  \centering
  \input{figures/hist_cf_report.tikz}
  \label{fig:phh_hist_sub1}
\end{subfigure}%
\begin{subfigure}[t]{.49\columnwidth}
  \centering
  \input{figures/hist_df_report.tikz}
  \label{fig:phh_hist_sub2}
\end{subfigure}
\caption{The separable distribution of the CF and DA Count features across labels in the noise-and-threshold phase of the vanilla PHH implementation.}
\label{fig:phh_histogram_vanilla_features}
\end{figure}

\para{Distinguishing attack via simple features.}
Assume the target device has either \texttt{URL0} or \texttt{URL1}. 
The attacker injects 99 Sybils with \texttt{URL0} before the target, so the input is either 100 copies of \texttt{URL0} or 99 copies of \texttt{URL0} and one \texttt{URL1}. 
The attacker obtains 1500 traces for each case and uses them to learn how to distinguish \texttt{URL0} and \texttt{URL1}. 
One leakage source in Figure~\ref{fig:vanilla-phh} is the noise-and-threshold loop iteration count, which depends on the number of keys in the input set. 

If the target visited \texttt{URL0}, there is just one key in the map, while a \texttt{URL1} visit results in 2 keys and therefore one more iteration. 
Figure~\ref{fig:phh_histogram_vanilla_features} highlights how we capture this leakage: 
The distributions of code fetches and data accesses in the noise-and-threshold phase are different across the two labels. 
This leakage can also be confirmed through a logistic regression classifier trained on the $\mathcal{F}_{1}$ feature set with perfect accuracy, giving the attacker a normalized advantage of 1.0 (Definition~\ref{def:dist-game}).

\para{Mitigation.} While padding the unordered map to the maximum number of possible keys is an obvious mitigation, it is extremely inefficient. 
Thus, we propose a DP-based mitigation. 
Observe that the noise-and-threshold loop iteration count---the quantity used by the previous attacker---has limited sensitivity to the target's value: switching from URL0 to URL1 changes it by only 1. 
If there is just enough variance in the number of iterations, the attacker will have a hard time distinguishing the two cases.
We accomplish this by injecting a random number of distinct ``dummy'' elements.
More precisely, we add a
number of dummies distributed as a discrete shifted Laplace random variable
with parameters $\epsilon, \delta$, to ensure that the number of loop iterations is $(\epsilon,\delta)$-DP and the corresponding leakage is bounded.  Figure~\ref{fig:phh-mitigation-noise} in Appendix \ref{appendix:phh} shows model code.
The idea of DP-fying side-channel leakage by adding dummy contributions appears in related work~\cite{sparsetwoservers, kobbistee}. We show that \FrameWorkName can help determine appropriate values for DP parameters.

We evaluate this mitigation by computing the empirical advantage of the attacker over the noise-and-threshold stage, across a range of $\epsilon$ values for the dummies, and comparing it with the analytical lower bound computed in Appendix~\ref{appendix:calculations} (Figure~\ref{fig:advantage_phh}, top). 
The results show that the mitigation succeeds for the noise-and-threshold phase, dropping the attacker advantage below and bringing it close to the analytical lower bound for sufficiently small $\epsilon$, across all feature sets described in Table~\ref{table:handcrafted_features}. 
This highlights that \FrameWorkName can evaluate defenses and also discover parameters for effective mitigations. Regarding performance, 
the expected number of dummies for $\delta = 10^{-9}, \epsilon = 0.1$ is about $200$, offering a very good tradeoff between privacy and performance for large enough deployments, e.g., with $n\geq 10000$.

\begin{figure}[t]
\setlength{\abovecaptionskip}{-5pt} 
\centering
\begin{subfigure}{\columnwidth}
  \centering
  \input{figures/advantage_phh_lr_report.tikz}
  \label{fig:sub1}
\end{subfigure}%
\hfill
\begin{subfigure}{\columnwidth}
  \centering
  \input{figures/advantage_phh_lr_accumulate.tikz}
  \label{fig:sub2}
\end{subfigure}
\caption{Advantage of the $\mathcal{F}_{1}$, $\mathcal{F}_{2}$, $\mathcal{F}_{3}$, $\mathcal{F}_{4}$ and $\mathcal{F}_{5}$ Distinguishing attacker for PHH protected by dummy operations for various $\epsilon$. 
The Empirical advantage is computed using a logistic regression for the noise-and-threshold (top) and Aggregate (bottom) stages, averaged over 5 trials. 
We compute the analytical upper bound using the formula in Appendix~\ref{appendix:calculations}.}
\label{fig:advantage_phh}
\end{figure}

\para{Distinguishing attack via advanced features.}
Despite the above mitigation, there could be leakage in the aggregation stage. 
In Figure~\ref{fig:advantage_phh} (bottom), we evaluate the mitigation against attacks that use \emph{all} feature sets over the aggregation subroutine. 
The input-dependent memory usage---code fetches and data accesses per page---expressed through $\mathcal{F}_{5}$ suffices for a logistic regression classifier to bypass the mitigation. 
The attacker bypasses the analytical upper bound because it is computed under the assumption that such features were hidden. Mitigations for this attack would need to randomize a histogram of accesses, rather than just one count. We leave this for future work.

\para{Fingerprinting attack.}
Recall that fingerprinting attacks are applicable when the attacker knows the distribution $W$ of a target's data over a (possibly large) domain and is interested in a (possibly small) set $I$. 
In our example, we instantiate $W$ with a power-law distribution with parameter $0.5$ over the list $L$ of the top 1000
most common sites in Alexa Top 1 Million Sites dataset~\cite{alexa-sites}.
The choice of $0.5$ is arbitrary but realistic, in that it corresponds to a skewed distribution.
The list $L$ contains URLs such as ``google.com'' and ``wikipedia.org'', but also URLs that might leak additional information, such as ``google.co.jp''. $I$ are the 301 URLs in $L$ not ending in ``.org'', ``.com'' or ``.net'', and thus often carrying information about the user's language or location.

The offline phase of the attack trains two classifiers: 
the membership {\em binary} classifier to identify whether or not the target's value is in $I$ and the fingerprint classifier to predict which of the $301$ values in $I$ the target had, assuming it had one. 
The offline phase has the following steps
\begin{compactenum}
    \item Create a dataset as follows:
        \begin{compactenum}
            \item Sample target's data $x$ from $W$.
            \item Create one instance of each element in $I$.
            \item To boost fingerprinting success, create enough Sybil data that is out-of-domain (e.g., not URLs) such that it is unlikely that the memory locations maintaining counts of elements in $I$ will be close to each other. This permits control over the granularity of side channels needed to effectively fingerprint $I$. 
            \item To improve membership detection in $I$, create enough Sybils such that a rehash event is guaranteed when $x\not\in I$. This can easily be done by inspecting the hashtable code, which is available to the attacker.
        \end{compactenum}
    \item Capture side-channel traces produced by running PHH algorithm on the above input.
    \item Repeat above to create sufficiently many unlabeled examples (traces).
    \item Train the membership classifier on \emph{all examples}, each labeled by the bit indicating if $x\in I$.
    \item Train the fingerprint classifier on \emph{only examples where $x\in I$}, each labeled by $x$.
\end{compactenum}

In the online phase, the attacker runs the membership classifier to determine whether or not the target has interesting data ($\in I$). If the target is deemed interesting, they will run the fingerprint classifier.

We evaluate the fingerprinting attack by building a dataset of 5000 instances, corresponding to traces of samples collected from $W$. 
We instantiate the membership and fingerprinting classifiers as logistic regression models based on all $\mathcal{F}_{1}$-$\mathcal{F}_{5}$ features. 
The membership classifier obtains a {\em normalized} advantage of 1.0 (Definition~\ref{def:dist-game}), which is a perfect score. 
This underscores the power of Sybils to exploit the vulnerability in the \texttt{std:unordered\_map} implementation. 
The fingerprinting classifier yields a {\em normalized} advantage of 0.44 (Definition~\ref{def:finger-game}), highlighting attack feasibility due to substantial leakage of the victim URL label through the PHH workflow. 

%% file: figures/hist_cf_report.tikz
\begin{tikzpicture}
\begin{axis}[
ybar,
style={font=\footnotesize},
xlabel={},
ylabel={},
width=\hsize,
height=2.5cm,
legend style={font=\scriptsize,at={(-0.3,-0.8)},anchor=west,draw=none,fill=none},
legend columns=2,
bar width=1.5,
]

\addplot+[mark size=0.5,blue!60,mark options={blue!60}, thick] table[x=Edges,y=Hist1,col sep=comma] {figures/hist_cf_report.csv};
\addplot+[mark size=0.5,orange!60, mark options={orange!60}, thick] table[x=Edges,y=Hist2,col sep=comma] {figures/hist_cf_report.csv};

\legend{lbl = 0 (CF),lbl = 1 (CF)};
\end{axis}
\end{tikzpicture}

%% file: figures/hist_df_report.tikz
\begin{tikzpicture}
\begin{axis}[
ybar,
style={font=\footnotesize},
xlabel={},
ylabel={},
width=\hsize,
height=2.5cm,
legend style={font=\scriptsize,at={(-0.3,-0.8)},anchor=west,draw=none,fill=none},
legend columns=2,
bar width=1.5,
]

\addplot+[mark size=0.5,blue!60,mark options={blue!60}, thick] table[x=Edges,y=Hist1,col sep=comma] {figures/hist_df_report.csv};
\addplot+[mark size=0.5,orange!60, mark options={orange!60}, thick] table[x=Edges,y=Hist2,col sep=comma] {figures/hist_df_report.csv};

\legend{lbl = 0 (DA),lbl = 1 (DA)};
\end{axis}
\end{tikzpicture}

%% file: figures/advantage_phh_lr_report.tikz
\begin{tikzpicture}
\begin{axis}[
style={font=\footnotesize},
xlabel={Epsilon},
ylabel={Advantage},
width=\hsize,
height=3cm,
legend style={font=\scriptsize,at={(0,1.3)},anchor=west,draw=none,fill=none},
legend columns=3,
xtick={0,1,2,3},
xticklabels={$\infty$,5,1,0.1}
]

\addplot+[dashed] table[x=x,y=analytical,col sep=comma] {figures/advantage_phh_lr_report.csv};
\addplot+ table[x=x,y=F1,col sep=comma] {figures/advantage_phh_lr_report.csv};
\addplot+ table[x=x,y=F2,col sep=comma] {figures/advantage_phh_lr_report.csv};
\addplot+ table[x=x,y=F3,col sep=comma] {figures/advantage_phh_lr_report.csv};
\addplot+ table[x=x,y=F4,col sep=comma] {figures/advantage_phh_lr_report.csv};
\addplot+ table[x=x,y=F5,col sep=comma] {figures/advantage_phh_lr_report.csv};

\legend{Analytical,LR F1,LR F2,LR F3,LR F4,LR F5};
\end{axis}
\end{tikzpicture}

%% file: figures/advantage_phh_lr_accumulate.tikz
\begin{tikzpicture}
\begin{axis}[
style={font=\footnotesize},
xlabel={Epsilon},
ylabel={Advantage},
width=\hsize,
height=3cm,
legend style={font=\scriptsize,at={(0,1.3)},anchor=west,draw=none,fill=none},
legend columns=3,
xtick={0,1,2,3},
xticklabels={$\infty$,5,1,0.1}
]

\addplot+[dashed] table[x=x,y=analytical,col sep=comma] {figures/advantage_phh_lr_accumulate.csv};
\addplot+ table[x=x,y=F1,col sep=comma] {figures/advantage_phh_lr_accumulate.csv};
\addplot+ table[x=x,y=F2,col sep=comma] {figures/advantage_phh_lr_accumulate.csv};
\addplot+ table[x=x,y=F3,col sep=comma] {figures/advantage_phh_lr_accumulate.csv};
\addplot+ table[x=x,y=F4,col sep=comma] {figures/advantage_phh_lr_accumulate.csv};
\addplot+ table[x=x,y=F5,col sep=comma] {figures/advantage_phh_lr_accumulate.csv};

\legend{Analytical,LR F1,LR F2,LR F3,LR F4,LR F5};
\end{axis}
\end{tikzpicture}

%% file: 08.udf.tex
\section{User-Defined Functions in Privacy Sandbox} \label{sec:udf}
TEE-isolated user-defined functions aim to enable new privacy-preserving applications such as data source verification~\cite{akkus2024praas}, outsourcing computation~\cite{goltzsche2019acctee}, private function-as-a-service~\cite{alder2019s}, and ads targeting~\cite{privacysandbox_fledge}.
UDFs run atop a language sandbox (e.g., Java\-Script/Wasm) to enable third-party queries on user data.
The language sandbox restricts third-party code from extracting user data, limits interfaces, and enforces constraints like time limits and accounting~\cite{goltzsche2019acctee}.
We focus on the side-channel evaluation of an example UDF that is implemented by the Protected Auction Key/Value service, part of the Privacy Sandbox~\cite{privacysandbox_udf}.
Privacy Sandbox, as an alternative to third-party cookies, enables third-party advertisers (AdTechs) to access advertising signals stored in an in-memory TEE-protected key/value database.
They use UDFs to run custom queries without direct access or logging. The query's output is aggregated and protected by DP techniques.

However, the requirement to keep the attacker out of the hypervisor requires special care for secure on-premise deployments of such privacy-preserving systems. 
Consider a hypothetical scenario where a malicious AdTech tries to run their own deployment of Privacy Sandbox on machines they fully control, including the hypervisor. 
Although the TEE would properly attest their key-value service deployment with the isolated UDF, they can still exploit side-channels to extract raw data from the key-value service.
Hence, specialized mitigations intended to prevent or detect side-channel signaling behavior by a UDF would be advisable.

\subsection{Stealing User Data via Covert Channel}
We evaluate a covert channel attack where a malicious UDF steals user query arguments.
Because the UDF is maliciously constructed, the attacker does not need to train a model to learn correlations. The UDF can use a deterministic encoding that gives a clear view of the data. As a result of this noiseless recovery, the normalized advantage is 1.0.

\para{Profiling UDF runtime.}
We analyze the UDF runtime to identify a trigger point for our covert-channel attack---when the receiver expects to see data from the sender. 
The UDF in the Protected Auction Key/Value service~\cite{privacysandbox_udf} is based on the V8 engine~\cite{v8}.
It supports JavaScript or inline Wasm, where the Wasm code must be invoked by JavaScript driver code, ensuring the UDF entry point remains in JavaScript.
When the V8 engine creates a typed array, it first executes one code page, writes to this page, and then executes this page again.
This \emph{X+W+X} access pattern serves as an indicator of the start and end of the UDF execution.
We mark the inline Wasm by surrounding it with two typed arrays, enabling \FrameWorkName\ to only track the Wasm execution.

\para{Encoding data over ciphertext.}
In the inline Wasm, we choose the ciphertext side channel to encode secret data. 
Specifically, we create a 4\,kB memory buffer that contains 256 ciphertext blocks.
As we iterate over the data stream byte by byte, we write to one of the 16-byte blocks in the 4\,kB buffer.
The index of the block depends on the value of each secret byte. 
We observe that in each iteration, the V8 engine consistently accesses four distinct memory pages in addition to our 4\,kB encoding buffer.
Among these, two page faults occur with a single specific block modification, which we attribute to updates in the loop counter and a temporary variable. 
The other two page faults are caused by memory reads.
Therefore, we check the ciphertext changes of the previous faulted page every time a new memory page fault occurs.
Thus, we can encode one byte with only five page faults, i.e., five context switches. 

\para{Evaluation.}
We evaluate the performance of our covert channel by transmitting 48 bytes of user-supplied input arguments 100 times.
We ignore three page faults when transferring each byte, as only one additional page fault is enough to trigger checking the ciphertext changes. 
Since these faulted pages have distinct page numbers, \FrameWorkName\ can simply skip unmapping them.
Our covert channel achieves an average transmission rate of 497 kbit/s with an error rate of 0.
The speed of this attack can be further increased by combining page numbers and ciphertext to encode secrets more efficiently.
For example, using 256 pages, an attacker can encode an extra byte per access through a controlled channel.
We verify this with a page fault-based covert channel spanning 256 pages.
To apply it, the attacker only needs a profiling step to map each page to its corresponding byte value, since the guest operating system controls the gPN, resulting in a non-contiguous mapping.
Given that the memory limit of the V8 engine is 4\,GB~\cite{privacysandbox_fledge}, this optimization meets such constraints. 

By enabling this practical exploit, \FrameWorkName reveals that simply placing the UDF in a language sandbox and restricting access to logging interfaces does not prevent the covert transfer of sensitive data across isolation boundaries.

%% file: 10.related.tex
\section{Related Work}
SGX-STEP~\cite{van2017sgx} is a framework for rapid prototyping of side-channel attacks in SGX~\cite{van2020lvi,constable2023aex}.
Similarly, Stacco~\cite{XiaoLCZ17} offers a framework for collecting and differentially analyzing multiple side-channel traces (e.g., page faults, cache attacks) against SGX enclaves. 
While Stacco focuses on using differential analysis to automatically detect vulnerabilities in cryptographic implementations such as SSL/TLS, our work aims to quantify information leakage in data-privacy applications. 
Within this evolving threat model, attackers use system interfaces to construct new side channels~\cite{xu2015controlled,wang2017leaky} and improve the reliability and bandwidth of side channels~\cite{moghimi2020copycat,van2017sgx,lee2017inferring}.
Unlike traditional side channels, these attacks can completely circumvent system noise.
We provide a more detailed discussion in Appendix~\ref{appendix:noise}.

SEV-SNP relies on encryption for hiding memory, but does not protect the ciphertext, which enables the new class of ciphertext side channels~\cite{li2021cipherleaks,li2022systematic}.
Attackers can also exploit privileged interfaces such as performance counters~\cite{gast2025counterseveillance} and power reporting~\cite{wang2023pwrleak} to leak side-channel information from CVMs.
TDXDown~\cite{wilke2024tdxdown} exploits gaps in system-level countermeasures against timer interrupt attacks.
SEV-STEP~\cite{wilke2023sev} is a framework for prototyping single-stepping and L1 cache attacks on SEV-SNP.

Software-based side-channel attacks have impacted TEEs in real products to steal cryptographic keys~\cite{ryan2019hardware,dall2018cachequote,XiaoLCZ17}.
The industry consensus to mitigate these attacks is to apply constant-time coding practices~\cite{corparation2021guidelines}.
Previous work has proposed automated tools to test such implementations~\cite{wichelmann2018microwalk,wang2017cached,bond2017vale,reparaz2017dude}.
However, these tools and constant-time coding practices are not applicable and practical for general-purpose programs.
Yuan et al. applied manifold learning to evaluate side-channel attacks on media software~\cite{yuan2022automated}.
Ciphertext side channels have been demonstrated as an effective technique to steal ML models' inputs and hyperparameters ~\cite{yuan2024ciphersteal,yuan2024hypertheft}.
Further, side-channel-assisted information retrieval has been demonstrated against SQLite~\cite{shahverdi2021database}.
We focus on automated side-channel testing of privacy-preserving applications in CVMs.

Haeberlen et al. argue that differentially private query release may be vulnerable to covert channel attacks via side-channel leakage \cite{HaeberlenPN11}. Their threat model explicitly separates the service provider from the adversary and leaves only the privacy budget and query time as side-channels. Jin et al. demonstrate that the running time of noise sampling algorithms could be used to circumvent DP guarantees \cite{JMRO21}. Ratliff \& Vadhan formalize DP against adversaries observing that side-channel and propose padding-based methods for achieving that objective \cite{ratliff-vadhan}.

%% file: 99.conclusion.tex
\section{Conclusion}
We conclude that automated analysis of side-channel leaks is crucial to improve the privacy guarantees of applications running within CVMs.
The status quo of relying solely on software techniques to mitigate side-channel attacks is impractical, and developers of privacy-preserving applications need to constantly evaluate an app's threat model and execution traces to ensure sufficient mitigation.
Toward this goal, a comprehensive framework like \FrameWorkName can significantly help developers assess their threat model and mitigation strategy.
In the future, defense-in-depth mitigations such as reducing side-channel information at the architecture level, preventing Sybil attacks, and carefully applying data-oblivious data structures like ORAM are promising but require further investigation.

%% file: 99a.ethics.tex
\section{Ethics Considerations}

This work builds upon the observation that Confidential Virtual Machines (CVMs) are not invulnerable to side-channel attacks. 
Although hardware vendors generally consider these attacks outside their threat models, practitioners deploying privacy-preserving solutions using CVMs are responsible for mitigating such attacks. 
Our proposed tool, \FrameWorkName, is designed to assist security researchers in identifying and mitigating side channels rather than serving as a tool for malicious exploitation. 
Although it could theoretically support attackers, our intent is to facilitate better defensive measures by quantifying risks and evaluating countermeasures.

We performed all experiments on local systems, did not use personal data or involve human subjects, and thus did not encounter any additional ethical concerns regarding privacy or user consent. 
By openly reporting vulnerabilities to the affected projects and contributing a framework for enhanced side-channel analysis, we aim to improve the overall security posture of CVM-based privacy solutions.


%% file: 100.appendices.tex
\subsection{Connecting DP \& Advantage}
\label{appendix:calculations}

\dpAdvantageLemma

Let $\rv{c}$ be the bit that indicates whether the target has $x_0$ or $x_1$. Observe that
\begin{align*}
\Pr[\attackeroutput_\attacker = \rv{c}] = \frac{1}{2}\big(& \Pr[\attackeroutput_\attacker = 0 ~|~\rv{c}=0] \\
&+ \Pr[\attackeroutput_\attacker = 1 ~|~\rv{c}=1] \big)
\end{align*}
because $\rv{c}$ is uniform over $\{0,1\}$.

Because $\Pr[\attackeroutput_\attacker = \rv{c}] > 1/2$, there must be at least one $i\in \{0,1\}$ where $\Pr[\attackeroutput_\attacker = i ~|~\rv{c}=i]>1/2$ otherwise the mean would be $\leq 1/2$. We rewrite the above equality using $i$:
\begin{align*}
&\Pr[\attackeroutput_\attacker = \rv{c}] \\
={}& \frac{1}{2}\big( \Pr[\attackeroutput_\attacker = i ~|~\rv{c}=i]\\
&+ \Pr[\attackeroutput_\attacker = 1-i ~|~\rv{c}=1-i] \big)\\
\leq{}& \frac{1}{2}\big( \Pr[\attackeroutput_\attacker = i ~|~\rv{c}=i]\\ &+e^\varepsilon \Pr[\attackeroutput_\attacker = 1-i ~|~\rv{c}=i] + \delta \big)
\end{align*}
The inequality comes directly from the definition of DP.

We can continue the analysis by adding and subtracting the quantity $\Pr[\attackeroutput_\attacker = 1-i ~|~\rv{c}=i]$:
\begin{align*}
=& \frac{1}{2}\big( \Pr[\attackeroutput_\attacker = i ~|~\rv{c}=i] + \Pr[\attackeroutput_\attacker = 1-i ~|~\rv{c}=i]\\
&+(e^\varepsilon-1) \Pr[\attackeroutput_\attacker = 1-i ~|~\rv{c}=i] + \delta \big)\\
=& \frac{1}{2}\big( 1+(e^\varepsilon-1) \Pr[\attackeroutput_\attacker = 1-i ~|~\rv{c}=i] + \delta \big)
\end{align*}
Now notice that $\Pr[\attackeroutput_\attacker = 1-i ~|~\rv{c}=i]$ is at most $1/2$ by virtue of the definition of $i$. Hence,
$$\Pr[\attackeroutput_\attacker = \rv{c}] \leq \frac{1}{2}\bigg( 1+ \frac{e^\varepsilon-1}{2} + \delta \bigg)$$
which in turn means that the advantage is $\leq \frac{e^\varepsilon-1}{4} + \frac{\delta}{2}$. We remark that this analysis is only useful for $\varepsilon< \ln (3-2\delta)$; otherwise, the advantage would be bounded by a number $\geq 1/2$.

\section{Mitigating Private Heavy Hitter Algorithm}
\label{appendix:phh}

\begin{figure}[h]
\begin{minipage}{0.45\textwidth}
\begin{minted}{c}
// batch, epsilon, and threshold are defined in 
// the context
std::unordered_map<string, int> hist;
std::vector<std::pair<string, int>> result;
...
// Mitigation params
double kEpsilonDummies = atof(argv[1]);
double kLaplaceScaleDummies = 2 / kEpsilonDummies;
double kTailBound = log(1 / kDelta) / kEpsilonDummies;
std::exponential_distribution<double> 
    expDummies(1 / kLaplaceScaleDummies);

// Aggregate inputs
...

// Add dummies
int nDummies = 0;
if (kEpsilonDummies > 0) {
    nDummies = 1 + 
        int(sample_centered_laplace(expDummies)) + 
        kTailBound;
}
for (int i = 1; i <= nDummies; ++i) {
    hist[-i]++;
}

// Noise and Threshold
...
}
\end{minted}
\end{minipage}
\caption{A mitigation for the obvious leakage in the Noise \& Threshold phase of the vanilla PHH implementation.}
\label{fig:phh-mitigation-noise}
\end{figure}

\Cref{fig:phh-mitigation-noise} shows our proposed DP-based mitigation.

\section{\FrameWorkName\ Implementation}
\label{appendix:framework_and_ml}

\subsection{Trace and Collection Speed}

\begin{figure}[ht!]
    \centering
    \includegraphics[width=\columnwidth]{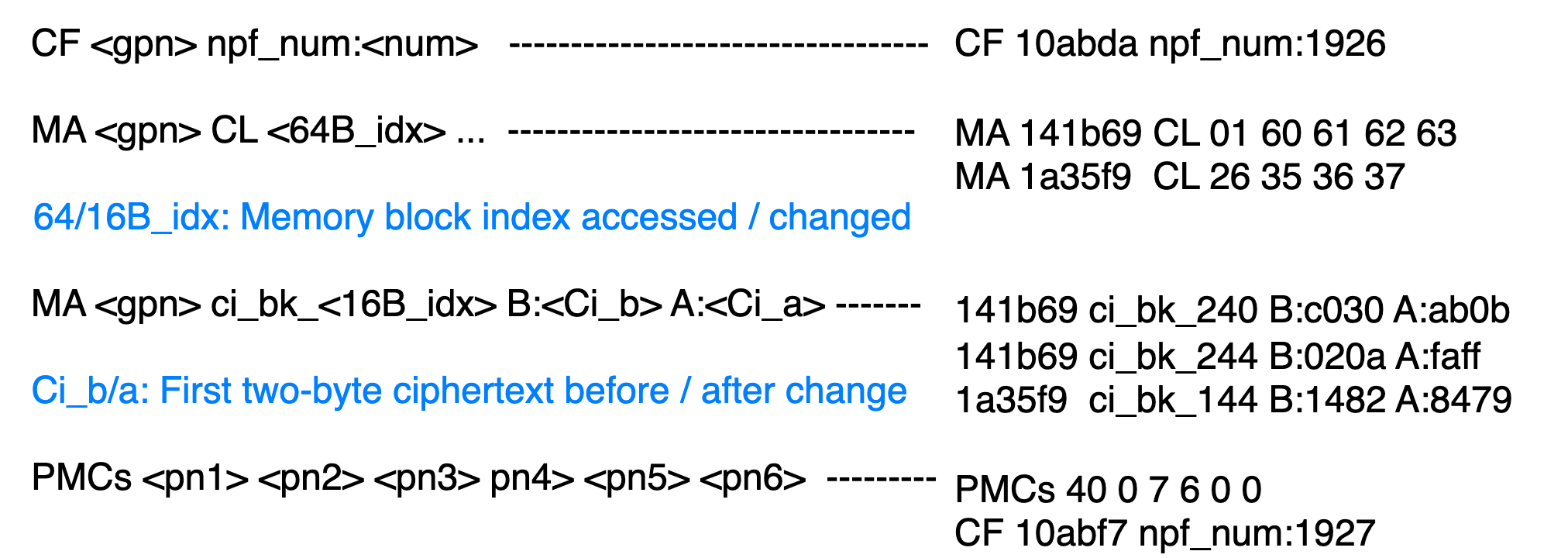}
    \caption{The syntax and example of side-channel trace. ``gpn'' represents the guest physical page number and ``num'' records the number of code pages monitored. ``pn'' represents the performance counter values of attacker-chosen events.}
    \label{fig:trace}
\end{figure}

\begin{table}[ht!]
\begin{center}
\resizebox{0.9\linewidth}{!}{
\begin{tabular}{ l|c|c}
\toprule
 & \textbf{Time per NPF} & \textbf{NPF Tracked/sec}  \\ 
\midrule
Page-level & 18,841 CPU cycles & 159,227  \\  \hline
Ciphertext & 22,543 CPU cycles & 133,079  \\  \hline
Cache attacks & 248,568 CPU cycles & 12,069 \\
\bottomrule
\end{tabular}}
\caption{The collection speed with different leakage choices with a $3.0$ GHz CPU.}
\label{table:col-speed}
\end{center}
\end{table}

\Cref{fig:trace} shows an example of the raw trace, where \textit{MA 141b69 CL 60} indicates access to the 60th 64B of the guest page at \texttt{0x141b69}.
\textit{ci\_bk} is followed by the 16B index in the page and the ciphertext value before and after the change.
\Cref{table:col-speed} shows the collection speed of \FrameWorkName\ under different leakage choices, averaged over 10,000 NPF using the \emph{Vanilla} application in \Cref{sec:pir} as the benchmark.
In cache attacks, we disable the hardware prefetcher and ensure a clean cache state at the prime state by executing the \texttt{wbinvd}\xspace instruction~\cite{zhang2024cachewarp}.

\begin{figure}[ht!]
    \centering
    \includegraphics[width=0.45\columnwidth]{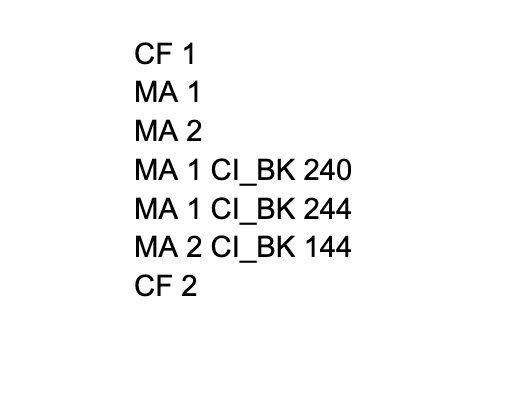}
    \caption{The trace in Figure~\ref{fig:trace} after pre-processing for sequence model.}
    \label{fig:trace_for_lstm}
\end{figure}

\para{Automatic Feature Learning}
To aid the sequence model, we pre-process the traces by abstracting the memory space, and maintaining information about the ciphertext. In Figure~\ref{fig:trace_for_lstm} we show the result of pre-processing the trace in Figure~\ref{fig:trace}.

\begin{table}[ht!]
\begin{center}
\begin{tabular}{ lll}
\toprule
\textbf{Layer} & \textbf{Dimm} & \textbf{\# Params} \\ \midrule
Input & 10,000 & 0 \\  \hline
Embedding & 64 & 640,000 \\  \hline
Bidirectional & 128 & 66,048 \\ \hline
Bidirectional & 64 & 41,216 \\ \hline
Dense & 64 & 4,160 \\ \hline
Dropout & 64 & 0 \\ \hline
Dense & 2 & 130 \\
 \bottomrule
\end{tabular}
\caption{The architecture of the LSTM model used throughout the case studies. }
\label{table:appendix_lstm}
\end{center}
\end{table}

\para{Sequence Model}
We implement an LSTM model according to the architecture described in Table~\ref{table:appendix_lstm}. The vocabulary size is set to 10,000, and the traces are truncated to the last 5000 tokens. The model uses a batch size of 32 and is trained for 50 epochs using early stopping.

\subsection{Intel CAT}
\label{appendix:intelcat}
We evaluate our optimized Prime+Probe attack on an Intel Xeon E5-2697 v4.
This platform features a 20-way inclusive L3 cache, where a "perfect" eviction set normally requires 22 (20+2) candidates mapping to the same slice and set~\cite{Rainer2025Rapid}. 
We employed the \texttt{pqos} utility to leverage Intel CAT, defining a new Class of Service (CLOS) with an L3 Capacity Bitmask (CBM) of \texttt{0xff}. 
This restricted assigned cores to 8 L3-cache ways (down from the default 20). 
We then associated the core executing the AES T-table benchmark with this restricted CLOS~\cite{Rainer2025RapidArtifacts}. 
This intervention yielded the expected outcome, reducing the required perfect eviction set size from 22 to just 10 (8+2) candidates. 
This finding confirms that a privileged attacker can co-opt hardware-based cache reservation mechanisms (e.g., Intel CAT or AMD MSRs) to force a victim into a shared, low-associativity partition.
This makes a successful Prime+Probe attack quantifiably less complex and resource-intensive.

\subsection{System Noise Circumvention}
\label{appendix:noise}

In this section, we discuss the noise sources for the four side-channel types and the hypervisor's ability to circumvente them.
\begin{itemize}
    \item  \textbf{Page Faults and Ciphertext:} Both of these channels are inherently deterministic. The sequence of page accesses is dictated by the victim program's execution flow, just as the ciphertext leakage is dictated by its memory write operations. Consequently, these channels are not susceptible to external system noise.
    \item  \textbf{PMCs:} PMCs are a per-core resource. A hypervisor can configure them to record only events originating from the guest VM. By scheduling other VMs (if any) to different cores, the hypervisor can dedicate a core to the victim, effectively eliminating any cross-VM interference and ensuring a noise-free PMC trace. 
    \item  \textbf{Cache Attacks:} The noise profile for cache attacks depends on the targeted cache level. We target the L2 cache, which is shared only within the physical core. Importantly, the shared L3 cache in the targeted AMD microarchitecture is non-inclusive. This means memory activities on other physical cores do not cause evictions in the L2 cache where the victim is sharing, thus preventing them from interfering with our L2-based Prime+Probe. The only remaining potential noise source is the sibling hyper-thread (SMT) on the same core. However, a privileged hypervisor can easily eliminate this by disabling SMT, granting the victim VM exclusive access to all L2 cache ways. 
\end{itemize}